\documentclass[12pt]{article}
\usepackage[usenames,dvipsnames,svgnames,table]{xcolor}
\usepackage{inputenc,fullpage}
\usepackage{graphicx}
\usepackage{amssymb}
\usepackage{kbordermatrix}
\renewcommand{\kbldelim}{(}
\renewcommand{\kbrdelim}{)}
\usepackage{pifont}
\newcommand{\xmark}{\text{\ding{55}}}%

\usepackage[margin=1.5in]{geometry}
\usepackage{showlabels}

\usepackage{changes}
\definechangesauthor[name={Ritwik}, color= ForestGreen]{Ritwik}
\setremarkmarkup{(#2)}

\usepackage{latexsym}
\usepackage{amsfonts}
\usepackage{mathrsfs}
\usepackage{verbatim}
\usepackage{setspace}
\usepackage{enumerate}
\usepackage{multirow}
\usepackage{booktabs}

\usepackage{amsmath}
\usepackage{amsthm}
\usepackage{subfig}
\usepackage{latexsym}
\usepackage{mathrsfs}
\usepackage{amsfonts}
\usepackage{color}
\usepackage{import}
\usepackage{appendix}
\usepackage{hyperref}
\usepackage{algorithm}
\usepackage{etex}
\hypersetup{
    unicode=false,          
    pdffitwindow=false,     
    pdfstartview={FitH},    
    pdftitle={My title},    
    pdfauthor={Author},     
    pdfsubject={Subject},   
    pdfcreator={Creator},   
    pdfproducer={Producer}, 
    pdfkeywords={keyword1} {key2} {key3}, 
    pdfnewwindow=true,      
    colorlinks=true,       
    linkcolor=blue,          
    citecolor=blue,        
    urlcolor=blue       
}
\RequirePackage[authoryear]{natbib}
\numberwithin{equation}{section}
\theoremstyle{plain}

\newtheorem*{theorem*}{Theorem}
\newtheorem{theorem}{Theorem}

\newtheorem{corollary}{Corollary}

\theoremstyle{remark}

\def\mathbold{\boldsymbol}

\def\bbeta{\boldsymbol{\beta}}
\def\bmu{\boldsymbol{\mu}}
\def\bgamma{\boldsymbol{\gamma}}
\def\bdelta{\boldsymbol{\delta}}
\def\btheta{\boldsymbol{\theta}}

\def\bbetak{\boldsymbol{\beta}_{(k)}}
\def\hbbetak{\widehat{\boldsymbol{\beta}}_{k}}

\def\btheta{\boldsymbol{\theta}}

\def\hbthetak{\widehat{\boldsymbol{\theta}}_{k}}

\def\bbetatrue{\bbeta_{\true}}
\def\bthetatrue{\btheta_{\true}}

\def\ellki{\ell_{k;i}}
\def\T{\mathsf{T}}

\def\tbbetak{\widetilde{\bbeta}_{k}}
\def\tbbetakstar{\widetilde{\bbeta}^{*}_{k}}
\def\bthetakstar{\btheta^{*}_{k}}

\def\by{\mathbold{y}}

\def\bX{\mathbold{X}}
\def\bx{\mathbold{x}}

\def\epsa{\varepsilon}
\def\bepsa{\mathbold{\varepsilon}}
\def\hbbeta{\widehat{\bbeta}}
\def\hmu{\widehat{\mu}}

\def\btheta{\mathbold{\theta}}
\def\hbtheta{\mathbold{\theta}}
\def\tbbeta{\bbeta_{k}}

\def\xstar{x^{*}}
\def\bSigma{\mathbold{\Sigma}}
\def\calN{\mathcal{N}}
\def\bzero{\mathbold{0}}
\def\bbE{\mathbb{E}}

\def\Re{\mathbb{R}}

\def\tell{\ell_{k}}

\def\telli{\ell_{k;i}}

\def\bbetak{\bbeta_{(k)}}

\def\hbbetak{\hbbeta_{(k)}}

\def\bbetak{\bbeta_{k}}
\def\bbetakstar{\bbeta_{k}^{*}}

\def\hbbetak{\hbbeta_{k}}

\def\tmatX{\widetilde{\matX}}

\def\bxstar{{\bx^{*}}}
\def\bxkstar{{\bx_{k}^{*}}}

\newcommand*{\Cdot}{\raisebox{-0.75ex}{\scalebox{1.8}{$\cdot$}}}

\newcommand{\bel}{\begin{eqnarray}\label}
\newcommand{\eel}{\end{eqnarray}}
\newcommand{\bes}{\begin{eqnarray*}}
\newcommand{\ees}{\end{eqnarray*}}

\def\convd{\stackrel{{\rm D}}{\longrightarrow}}
\def\toD{\convd}

\def\benu{\begin{enumerate}}
\def\eenu{\end{enumerate}}

\def\argmax{\mathop{\rm arg\, max}}
\def\argmin{\mathop{\rm arg\, min}}

\def\Re{{\mathbb{R}}}

\def\complex{\mathop{{\rm I}\kern-.58em\hbox{\rm C}}\nolimits}
\def\pa{\partial}

\def\diag{\hbox{diag}}

\def\rank{\hbox{rank}}

\def\mathbold{\boldsymbol} 

\def\bb{\mathbold{b}}

\def\matB{{\rm \textbf{B}}}

\def\bc{\mathbold{c}}

\def\bbE{\mathbb{E}}

\def\hbar{\overline{h}}
\def\matH{{\rm \textbf{H}}}

\def\matI{{\rm \textbf{I}}} \def\bbI{\mathbb{I}}

\def\matJ{{\rm \textbf{J}}}

\def\bell{\mathbold{\ell}}

\def\calM{{\cal M}}

\def\calN{{\cal N}}

\def\sfN{\mathsf{N}}

\def\calO{{\cal O}}

\def\bp{\mathbold{p}}\def\phat{\widehat{p}}
\def\hbp{{\widehat{\bp}}}

\def\bbP{\mathbb{P}}

\def\bu{\mathbold{u}}

\def\Ubar{{\overline U}}

\def\Vbar{{\overline V}}

\def\bw{\mathbold{w}}

\def\matW{{\rm \textbf{W}}}

\def\bx{\mathbold{x}}

\def\bX{\mathbold{X}}

\def\matX{{\rm \textbf{X}}}

\def\by{\mathbold{y}}

\def\bzero{\mathbold{0}}

\def\bbeta{\mathbold{\beta}}

\def\hbbeta{{\widehat{\bbeta}}}\def\tbbeta{{\widetilde{\bbeta}}}
\def\bbetak{\mathbold{\beta}_{k}}
\def\hbbetak{\widehat{\bbeta}_{k}}

\def\bgamma{\mathbold{\gamma}}

\def\hbgamma{{\widehat{\bgamma}}}

\def\bdelta{\mathbold{\delta}}

\def\vepsa{\varepsilon}\def\epsa{\epsilon}
\def\bepsa{\mathbold{\vepsa}}

\def\btheta{\mathbold{\theta}}

\def\hbtheta{{\widehat{\btheta}}}

\def\bmu{\mathbold{\mu}}\def\hmu{\widehat{\mu}}

\def\hbmu{{\widehat{\bmu}}}

\def\hsigma{\widehat{\sigma}}

\def\bSigma{\mathbold{\Sigma}}

\newcommand{\norm}[2]{\|#1\|_{#2}}

\def\boxit#1{\vbox{\hrule\hbox{\vrule\kern6pt
\vbox{\kern6pt#1\kern6pt}\kern6pt\vrule}\hrule}}
\def\hlcomment#1{\vskip 2mm\boxit{\vskip 2mm{\color{blue}\bf#1} {\color{green}\sf \hfill--HL\vskip 2mm}}\vskip 2mm}

\newcommand{\blind}{1}
\allowdisplaybreaks
\def\drop{\mbox{\rm\tiny drop}}
\def\true{\mbox{\rm\tiny true}}

\usepackage{JASA_manu}

\begin{document}

\title{\large{A GENERAL FRAMEWORK FOR FREQUENTIST MODEL AVERAGING }\if1\blind{\footnote{ \baselineskip=12pt
Corresponding Author: {Min-ge Xie (mxie@stat.rutgers.edu)}.
This article is a work developed based on the thesis of the first author.
The research was supported in part by US NSF grants DMS-1513483 (MX), DMS-1418042 (HL), and by Award No.11529101(HL), made by National Natural Science Foundation of China.
}}
\fi
}
\if1\blind
\author{Priyam Mitra$^*$, Heng Lian$^\dagger$, Ritwik Mitra$^{\dagger\dagger}$, Hua Liang$^\ddagger$ and Min-ge Xie$^*$  \\
Rutgers University$^*$, City University of Hong Kong$^\dagger$, \\ Princeton University$^{\dagger\dagger}$ \&  George Washington University$^\ddagger$ }
\fi

\date{}
\maketitle

\begin{center}
\textbf{Abstract}
\end{center}

{ \singlespace
Model selection strategies have been routinely employed to determine a model for data analysis in statistic, and further study and inference then often proceed as though the selected model were the true model that were known a priori. 
This practice does not account for the uncertainty introduced by the selection process and the fact that the selected model can possibly be a wrong one. Model averaging approaches try to remedy this issue by combining estimators for a set of candidate models. 
Specifically, instead of deciding which model is the `right' one, a model averaging approach suggests to fit a set of candidate models and average over the estimators using certain data adaptive weights. In this paper we establish a general frequentist model averaging framework that does not set any restrictions on the set of candidate models. It greatly broadens the scope of the existing methodologies under the frequentist model averaging development. Assuming the data is from an unknown model, we derive the model averaging estimator and study its limiting distributions and related predictions while taking possible modeling biases into account. We propose a set of optimal weights to combine the individual estimators so that the expected mean squared error of the average estimator is minimized. Simulation studies are conducted to compare the performance of the estimator with that of the existing methods. The results show the benefits of the proposed approach over traditional model selection approaches as well as existing model averaging methods.

}
\vspace*{.1in}

\noindent\textsc{Keywords}: {Asymptotic distribution; Bias variance trade-off; Local mis-specification; model averaging estimators; Optimal weight selection. }


\newpage

\section{Introduction}

When there are several plausible models to choose from but no definite scientific rationale to dictate which one should be used, a model selection method has been used traditionally to determine a `correct' model for data analysis. Commonly used model selection methods, such as Akaike information criterion (AIC), Bayesian information criterion (BIC), stepwise regression, best subset selection, penalised regression, etc., are data driven and different methods may use different criteria; cf., e.g., \cite{Hastie2009} and reference therein.
Once a model is chosen, further analysis proceeds as if the model selected is the true one. This practice does not account for the uncertainty introduced in the process due to model selection, and can often lead to faulty inference as discussed in \cite{Madigan1994, Draper1995, Buckland1997}, among others. To provide a solution to the problem, model averaging methods have been introduced to incorporate model uncertainty during analysis; cf., e.g., \cite{Claeskens2008}. Instead of deciding which model is the `correct' one, a model averaging method uses a set of plausible candidate models and final measures of inference are derived from a combination of all models. The candidate models are combined using some data-dependent weights to reflect the degree to which each candidate model is trusted.


Our research on model averaging is motivated in part by a real life example on a prostate cancer study where the relationship between the level of prostate-specific antigen and a number of clinical measures in men who were about to receive a radical prostatectomy was investigated.
The variables included in the study are log cancer volume, log prostate weight,
age, log of the amount of benign prostatic hyperplasia, seminal vesicle invasion, log of capsular penetration, Gleason score, and percent of Gleason scores 4 or 5. In analysis of such data, 
a common theme is that different model selection methods may choose different models as the `true' one. For example, AIC and BIC, two commonly used model selection criteria, may pick two different models, as the criteria for selection is different. Such situations would certainly raise many questions in practice. For instance,
if the estimator is selected by using a model selection criteria, how would we address the possibility
that the selection is a wrong model? Also, if different model selection methods give us different results, we might wonder how trustworthy the model selection procedures are.
Instead of choosing one model using a model selection scheme, we can use an average of estimators from different models. The model averaging estimator then can provide us with an estimate of any parameter involved in the study and can be used for providing confidence bounds. The model averaging estimator can be used for prediction purposes as well.

\cite{Hjort2003} provided a formal theoretical treatment of frequentist model averaging approaches, which provided in-depth understanding of model averaging approaches and was well cited.
However, the development had an assumption that any extra parameters not included in the narrowest model will shrink to zero at a $ \calO({1}/{\sqrt n}) $ rate.
It essentially requires that the all candidate models are within a $ \calO({1}/{\sqrt n}) $ neighborhood of the true model. Although this assumption avoids a technical difficulty of handling biased estimators, in reality we do not know the true model and thus excluding from consideration those models that are beyond $ \calO({1}/{\sqrt n}) $ neighborhood of the true model appears to be very restrictive in practice.
In this paper, we remove this restrictive assumption in \cite{Hjort2003} and develop frequentist model averaging approaches under a much more general framework. Our model averaging scheme allows us to use all the potential candidate models available, even the ones that produce biased estimates.

The development is motivated by the familiar bias-variance trade-off. If we use an overly simple model, the parameter estimates will often be biased, but it can also possibly have less variance, because there are fewer parameters to estimate. Similarly, if a bigger model is used, the parameter estimates often have low or no bias but increased variance. It is possible that biased estimators may end up having lower mean squared error (MSE) than the bigger model or even true model, and visa verse. In our development, we study the
delicate
balance between bias and variance in all possible models and utilize the knowledge to develop new frequentist model averaging approaches.

 
A key element of a model averaging method is selection of weights that help us build a combined model averaging estimator. The weights proposed in our development are based on the aforementioned bias-variance trade-off, anchoring on the mean squared error (MSE) of the overall model averaging estimator. The weighing scheme is similar to but not the same as that discussed in \cite{Liang2011}, in which the authors only focused on  Gaussian linear regression models. Specifically, a consistent estimate of the mean squared error of the model averaging estimator is proposed, and the weights are chosen such that the MSE estimate is minimized. Using these weights, we show that model averaging performs better or no worse than several existing and commonly used model selection or model averaging methods. In particular,  the weights chosen often display good optimality properties, for example, the parameter estimates converging to the true parameter values as sample size $n$ goes to infinity. Thus it can be shown that in most of the cases weights that are chosen to combine the candidate models highlight the contribution of the true model. However, for a finite sample size with $n$, biased estimators may end up having lower mean squared error than that from the true model and the model averaging estimator may be based on biased candidate models.

A model averaging estimator incorporates model uncertainty into the analysis by combining a
set of competing candidate models rather than choosing just one. It also provides an insurance against selecting a poor model thus improving the risk in
estimation.
In \cite{Hjort2006} and \cite{Claeskens2008}, variable selection methods for the Cox proportional hazards regression model were discussed along with the choice of weights. In \cite{Hansen2007} a new set of weights was derived using Mallow's criterion. In \cite{Liang2011}, the authors proposed an unbiased estimator of the risk and a set of optimal weights was chosen by minimizing the trace of the unbiased estimator. Further details about model selection and averaging can also be found in \cite{Lien2005,Karagrigoriou2009,wan.zhang.ea:2010,zhang.wan.zhou:2012, Wei2012}. 
The model averaging method has also been  used in many areas of applications, e.g.,
\cite{Danilov2004a, Danilov2004} for forecasting stock market data,
\cite{Pesaran2009} for risk of using false models in portfolio management, \cite{magnus.wan.ea:2011} for analysis of
the Hong Kong housing market, and \cite{Posada2004} for a study of
phylogenetics in biology. Our development in this article extends the existing theoretical frequentist development to a general framework so it can incorporate biased models under a general setting.  
Model averaging has been also discussed in the Bayesian framework; see, e.g. \cite{Raftery1998} and \cite{ Hoeting1999a}. In a Bayesian approach, a weighted average of the posterior distributions under every available candidate model was used for estimation and prediction purposes. The weights were determined by posterior model probabilities. Model averaging in a frequentist setup, as in \cite{Hjort2003} and also ours, precludes the need to specify any prior distributions, thus removing any possible oversight due to faulty choice of priors. The question in a frequentist setting is how to obtain the weights by a data-driven approach. 

The rest of the article is organized as follows. In section \ref{sec:2}, we propose a general framework that covers the framework of \cite{Hjort2003} as a special case and study asymptotic properties of model averaging estimators. We also derive a consistent estimator for the mean squared error of the model averaging estimator and use it to facilitate our choice of data-driven weights in section \ref{sec:3}.  The development is illustrated in generalized linear models and particularly in linear and logistic model setups. In section \ref{sec:4}, simulation studies are carried out to examine the performance of the proposed estimator and to compare its performance with existing methods.

\section{General Framework}\label{sec:2}
\subsection{Notations and Set up}\label{subsec:notation}

Consider $n$ independent data points $\by=(y_{1}, \cdots, y_{n})$ sampled from a distribution having density of the form $f(y)\equiv f(y,\bbeta)$,  where $\bbeta$ is the unknown parameter of interest.
Here the parameter $\bbeta$ can be written as $\bbeta=(\btheta,\bgamma)$, where $\btheta \in \Theta \subset \Re^{p}$, $p \ge 0$, are the parameters that are certainly included in every candidate model and $\bgamma \in \Re^{q}$ is the remaining set of parameters that may or may not be included in the candidate models. We assume that $p$ and $q$ are given.
As a model averaging method, instead of choosing one particular candidate model as the ``correct'' model, we consider a set of candidate models, say $\calM$, in which each candidate model contains the common parameters $\btheta$ and a unique $\bgamma'$ that includes $m$ of $q$ components of the parameter $\bgamma$, $0 \leq m \leq q$.

The choice of $\calM$ can vary depending on the problem that one is trying to solve. For example, the candidate model set $\calM$ can contain all possible $2^{q}$ combinations of $\bgamma$. Or, one can choose a subset of the $2^{q}$ possible models as $\calM$. In \cite{Hansen2007}, a set of nested models has been used as candidate models, with $|\calM|=q+1$. In \cite{Hjort2003} $\calM$ includes candidate models that are within a $ \calO({1}/{\sqrt n}) $ neighborhood of the true model. Our development encompasses both setups as there are no restrictions on $\calM$, and $\calM$ can include any number of candidate models between $1$ and $2^q$. Similar setup was used in \cite{Liang2011} where $\calM$ is also unrestricted, but the development there was done in the standard linear regression framework.

Let the parameters in the true model be given by $\bbetatrue = (\btheta_{\true}, \bgamma_{\true})$. Let $m^{\true}$ be the number of components of $\bgamma$ that are present in the true model. Define $\calM_{\in} $ as the collection of the candidate models that contain the true model, thus every model in $\calM_{\in}$ contain each and every one of the $m^{\true}$ components of $\bgamma$.
Define $\mathcal{M}_{\notin}= \calM- \calM_{\in}\subset \calM$, so $\mathcal{M}_{\notin}$ contains candidate models for which at least one of those $m^{\true}$ components are not present. Clearly  $\mathcal{M} = \mathcal{M}_{\in} \cup \mathcal{M}_{\notin} $.

In \cite{Hjort2003},  
a common parameter is also present in all the candidate models that is similar to ours. But the treatment of $\bgamma$ is different. In particular, the model containing just $\btheta$ is called a narrow model and the true model is chosen of the form $f(\by)=f(\by,\btheta ,\bgamma_0 + \delta/\sqrt n)$. Here, parameter $\delta$ determines how far a candidate model can vary from the narrow model and $\bgamma_0 $ is a given value of $\bgamma$ for which any extended model reduces down to the narrow model.
Thus, this choice of true model essentially requires that the all candidate models are within a $ \calO({1}/{\sqrt n}) $ neighborhood of the true model.  Any model that is beyond $ \calO({1}/{\sqrt n}) $ neighborhood of the true model is excluded from the analysis. In this paper, we remove this rather restrictive constraint. Indeed, we assume the parameter for the true model is $\bbetatrue = (\btheta_{\true}, \bgamma_{\true})$, where $\bgamma_{\true}$ may or may not have any of the $q$ components, and the candidate model set $\calM$ can be a subset or contain all possible $2^{q}$ combinations of $\gamma$. Thus in our model setup there are no restrictions on the choice of true model or on the set of candidate models as in \cite{Hjort2003}. Furthermore, we can treat the setup considered in \cite{Hjort2003} as a special case of ours by restricting $\bgamma_{\true}$ so that all the candidate models will have a bias of order $\calO({1}/{\sqrt n})$ or less.



Note that, every candidate model includes a unique $\bgamma$ that may or may not include all $q$ components. Thus the numbers of parameters from different candidate models may be different.
For ease of presentation and following \cite{Hansen2007}, we introduce an augmentation scheme to bring all of them to the same length. We first illustrate the idea using the regression example considered by \cite{Hansen2007}: $\by$ is the vector of responses, $\bX$ is the design matrix with full column rank $p+q$ and the candidate models are nested models. We further assume the first $p$ columns of $\bX$ are always included in the candidate models; the special case with $p =0$ goes back to the setup of \cite{Hansen2007}. It follows that the $k^{\rm th}$ candidate model includes the first $p+k$ columns of $\bX$, $k=0,\cdots,q$. Denote by $\hbbeta_{k}$ the estimated regression parameters corresponding to the $k^{\rm th}$ candidate model. Then the $(p+k) \times 1$ vector $\hbbeta_{k}$ can be augmented to a $(p+q) \times 1$ vector $(\hbbeta_{k}^{\top}, {\bf 0}^{\top})^{\top}$, by adding $(q - k)$ $0$'s. The augmented estimator for the $k^{\rm th}$ candidate model is given by
\begin{equation}
\tbbeta_{k}= (\hbbeta_{k}^{\top}, {\bf 0}^{\top})^{\top} =\begin{bmatrix}
(\bX_{k}^{\top}\bX_{k})^{-1} \bX_{k}^{\top}\by\\
0
\end{bmatrix};
\label{eq:augmented}
\end{equation}
cf., e.g., \cite{Hansen2007} which adopted this augmentation on a set of nested candidate models. 

More generally, let $\bbetak$ be the parameter for the $k^{{\rm th}}$ model in $\calM$. Assume the length of $\bbetak$ is $p+m_k$, where $m_k$ depends on $k$. Define the log-likelihood for the $i^{{\rm th}}$ observation in the $k^{{\rm th}}$ model as $\ell_{k;i}(\bbeta_{k}) = \log f(y_{i}, \bbeta_{k})$. The maximum likelihood estimate (MLE) of $\bbetak$ using the $k^{{\rm th}}$ model is $\hbbeta_{k} = \argmax_{\beta_k} \ell_{k}(\bbetak)$, where  $\ell_{k}(\bbetak) = \sum^{n}_{i=1}\ell_{k;i}(\bbeta_{k})$.
Write the score function of the $k^{{\rm th}}$ model as $S_{k}(\bbeta)$.
As in the example above, the vector $\bbeta_{k}$ for the $k^{{\rm th}}$ model can be augmented to a $(p+q) \times 1$ vector 
$(\bbeta_{k}^{\top}, \bc_k^{\top})^{\top}$, where $\bc_k$ is a fixed value used for augmentation to hold spaces. 
The augmented maximum likelihood estimator is given by
$\tbbeta_{k} = (\hbbeta_{k}^{\top}, \bc_k^{\top})^{\top}$.
The fixed value augmentation does not affect the parameter, and only appends the length of the parameter. In the linear model example above the values $\bc_k=0$.
Some examples of $\bc_k \not =0$ can be found in \cite{MitraThesis}.
Similarly, $\bbetatrue$ can be augmented to a $(p+q) \times 1$ vector $(\bbetatrue^{\top}, \bc^{\top})^{\top}$ for a certain fixed set of $\bc$ without altering the true model. Thus, without loss of generality and from now on, we assume $\bbetatrue$ is a $(p+q) \times 1$ vector in the sense that some of the elements may be the augmented to fill the space.

For the model $k \in \calM$, let us define $\bbetakstar \in \Re^{p+m_k}$ as the solution of the equation $\bbE S_{k}(\bbeta)=0$, where $S_{k}(\bbeta)$ is the score function of the $k^{{\rm th}}$ model having $p+m_k$ parameters. Define, as before, $\tbbetakstar \in \Re^{p+q}$ as the $\bc-$augmented version of $\bbetakstar$. Since the score function is Fisher consistent,  $\tbbetak \rightarrow \tbbetakstar$  under usual regularity conditions. But this
 $\tbbetakstar$ may not be close~to~$\bbetatrue$.

Let $\bmu: \Re^{p+q}\rightarrow \Re^{\ell}$ be a  general function that is $1^{{\rm st}}$ order partially differentiable and $\bmu$ $ = $ $\bmu (\bbeta_{\true})$ is the parameter of interest.
Then, the model averaging estimator of $\bmu$ is defined~as
\begin{align}\label{fmae} 
\widehat\bmu_{ave} = \sum_{k \in \mathcal{M}}  \ w_k \bmu ( \tbbeta_k ),
\end{align}
where the weights $0\leq w_{k}\leq 1,  \forall\ k$, and $\sum_{k \in \calM}{w_{k}}=1$. 
In the remainder of this section, we derive the asymptotic properties of the model averaging estimator (\ref{fmae}) for any given set of weights~$w_{k}$.



\subsection{Main Results}

We assume the usual regularity  conditions under which the familiar likelihood asymptotic arguments apply; cf.,  the conditions listed in the Appendix. See also \cite{Lehmann1998, Lehmann1999, van00} for more details.


Let 
$\nabla\bmu \in \Re^{\ell\times (p+q)}$ be the first order derivative of the $\Re^{p+q}\rightarrow \Re^{\ell}$ function $\bmu$.
Define $\mathbf{H}_{k}$= $\lim_{n\rightarrow \infty} \dfrac{1}{n} \bbE\left[\tell''(\bbetakstar)\right]$ and  assume it is invertible. We also assume
\begin{align*}\label{eq:lindcond1}
{\rm (\textbf{A}1)} \,\, \lim_{n} \dfrac{1}{n}\sum^{n}_{i=1} \bbE \left[\max_{k \in \calM} \norm{\nabla\bmu(\tbbetakstar)\matH_{k}^{-1}\ellki'(\bbetakstar)}{}\ \bbI \ \left\{\max_{k \in \calM} \norm{\nabla\bmu(\tbbetakstar)\matH_{k}^{-1}\ellki'(\bbetakstar)}{}>\sqrt{n}\epsa\right\}\right] = 0,
\end{align*}
for any $\epsa>0$, where $\bbI \{\cdot\}$ is the indicator function.
We have the following theorem. Its proof is given in the Appendix.

\begin{theorem} \label{th:main2} 
Let $\tbbeta_{k}$ be the $\bc$-augmented MLE as defined in \eqref{eq:augmented} for the $k^{\rm th}$ model in $\calM$. Let $0\leq w_{k}\leq 1$ for $k \in \calM$ be model weights so that $\sum_{k}w_{k}=1$. Assume condition $(\textbf{A}1)$ holds. The asymptotic distribution of the model averaging estimator for $\mathbf{\bmu}(\bbetatrue)$ is given as,
\begin{equation} \label{eq:mu}
\sqrt{n}\sum_{k \in \calM} w_{k}\{\bmu(\tbbetak) - \bmu(\bbetatrue)\} - \sqrt{n}\sum_{k \in \calM} w_{k}(\bmu(\tbbetakstar)-\bmu(\bbeta_{\true})) \toD  \calN\left(0, \ \bSigma_{w}\right),
\end{equation}
where the variance $\bSigma_{w}$ is given by 
\begin{align} \label{eq:varmu}
\bSigma_{w} =  \lim_{n\rightarrow \infty} \dfrac{1}{n}\sum^{n}_{i =1} \bbE \left[ \big(\sum_{k}w_{k}\nabla\bmu(\tbbetakstar)^{\top}\matH^{-1}_{k}\ellki'\big)^{\otimes 2}\right]
\end{align}
\end{theorem}

The condition (\textbf{A}1) implies that the contribution of $\nabla\bmu(\tbbetakstar)\matH_{k}^{-1}\ellki'(\bbetakstar)$
to the total variance, for each model $k$ in the set $\calM$ and for each $1 \leq i \leq n$ is asymptotically negligible, and it is satisfied in a wide array of cases. We provide a set of sufficient conditions under which it is satisfied, and we also provide such examples in the cases of linear and generalized linear models in Section \ref{sec:3}.
 See further discussions of the condition in Section \ref{sec:3}.

In our general framework, there is no guarantee that  $\tbbetakstar = \bbetatrue$, neither does $||\tbbetakstar - \bbetatrue|| \to 0$ asymptotically, particularly when $k \in \mathcal{M}_{\notin}$. So $\bmu(\tbbetakstar) - \bmu(\bbetatrue)$ is not necessarily $0$, even asymptotically. But we can view it as a measurement of the bias by the $k^{\rm th}$ model. Thus, with the second term on the left hand side of (\ref{eq:mu}) serving as a bias correction term,
Theorem~\ref{th:main2} states that the model averaging estimator still retains the usual form of asymptotic normality after the bias correction.
In the theorem, the weights are fixed. In practice, we often estimate the weights using data. In this case, we need that the estimated weight $w^{(n)}_k(\by)$ for the $k$th model converges to $w_k$ as $n$ goes to infinity. By Slutsky's lemma, the result in Theorem \ref{th:main2} still holds. A further study of data dependent weights is in Section~\ref{sec:3}.

All the candidate models have $\btheta$ in common. We can use Theorem \ref{th:main2} to construct asymptotic convergence results for the common parameter $\btheta$. If we consider a function from $(\btheta,\bgamma) \mapsto \btheta$ to extract the $\btheta$ parameter, then by a direct application of Theorem \ref{th:main2} we can derive the asymptotic distribution of $\btheta$ as given below in Corollary \ref{cor:theta}.

\begin{corollary} \label{cor:theta}
Let $\btheta$ be the common parameter for all candidate models in $\calM$. Let $\bbetatrue=(\bthetatrue,\bgamma_{\true})$, $\bbetakstar=(\bthetakstar,\bgamma^{*}_{k})$, $\hbbeta_{k}=(\hbtheta_{k},\hbgamma_{k})$. Then under the same setup as in Theorem \ref{th:main2}
\begin{align}
\sqrt{n}\sum_{k \in \calM} w_{k}(\hbthetak - \bthetatrue) - \sqrt{n} \sum_{k \in \calM} w_{k}(\bthetakstar - \bthetatrue)  \toD \calN\left(0, \ \bSigma_{w}\right),
\end{align}
where the variance is given by
$\bSigma_{w} =  \lim_{n\rightarrow \infty} \dfrac{1}{n}\sum^{n}_{i =1} \bbE \left[ \big(\sum_{k}w_{k} [\matI_p, \bzero]\matH^{-1}_{k}\ellki'\big)^{\otimes 2}\right]$. 
\end{corollary}

\subsection{Connection to \cite{Hjort2003}'s development}\label{sec:3Hjort}
The development of \cite{Hjort2003} required that all candidate models are within a $\calO({1}/{\sqrt n}) $ neighborhood of the true model. We broaden this framework in our development. 
In particular, we show in this subsection that the results described in \cite{Hjort2003} can be obtained as a special case of our result.

We start with a description of the misspecified model setup used in \cite{Hjort2003}. Let $ Y_1, \cdots, Y_n $ be a independent and identically distributed (i.i.d.) sample from density
$f$ of maximum $p+q$ parameters. The parameter of interest is $ \mu = \mu(f)$, where $\mu :\Re^{p+q} \rightarrow \Re$ .
The model that includes just $p$ parameters, say $\btheta$, is defined as the narrow model, while any extended model $f(\by,\btheta,\bgamma)$ reduces to the narrow model for $\bgamma=\bgamma_0$; here the vector $\bgamma_0$ is fixed and known. For the $k^{{\rm th}}$ model with unknown parameters $(\btheta,\bgamma_k)$, the MLE of $\mu$ is written as ${\hmu_k} = \mu({\hbtheta_k},{\hbgamma_k},\bgamma_{0,k^c})$, where $k^c$ refers to the elements that are not contained in ${\bgamma_k}$.
Thus in this setup, if a parameter $\gamma_j$ is not included in the candidate model, we set $\gamma_j=\gamma_{j,0}$, the $j^{\rm th}$ element of $\bgamma_0$. The true model is assumed to be
\begin{align}\label{eq:mispec}
f_{\true}(y)=f(y,\btheta_{0},\bgamma_0+\bdelta/\sqrt{n}),
\end{align}
where $\bdelta$ signify the deviation of the model in directions $1, . . . , q$. So $\bbetatrue= (\btheta_{0},\bgamma_0+\bdelta/\sqrt{n})$. Let us write $\bbeta_{0} = (\btheta_{0},\bgamma_0)$. We will also write $\mu_{\true} = \mu(\bbetatrue)$, which is the estimand of interest. Under this model setup, \cite{Hjort2003} derived asymptotic normality result for the model averaging estimator $\sum_{k}w_{k}\hmu_{k}$. To describe their result, let us first define
\begin{align*}
S(y) = \begin{bmatrix}
U(y)\\
V(y)
\end{bmatrix}
=
\begin{bmatrix}
{\partial \over \partial\btheta} \log f(y,\btheta,\bgamma)\\
{\partial \over \partial\bgamma} \log f(y,\btheta,\bgamma)
\end{bmatrix} \bigg|_{\btheta =\btheta_0,\bgamma =\bgamma_0} \text{ and }{\rm var}\{S(Y)\} =
\begin{bmatrix}
\matJ_{00} \quad \matJ_{01}\\
\matJ_{01} \quad \matJ_{11}
\end{bmatrix} = \matJ_{full}, \text{ say}.
\end{align*}
Let $\Ubar_{n}= n^{-1}\sum_{i}U(Y_{i})$ and $\Vbar_{n} = n^{-1}\sum_{i}V(Y_{i})$. Denote by $V_{k}(Y)$ and $\Vbar_{n;k}$ the appropriately subsetted vectors obtained from $V(Y)$ and $\Vbar_{n}$, with the subset indices corresponding to that of $\hbgamma$  in model $k \in \calM$,  respectively. Also, define $\matJ_{k} = \mbox{\rm var}\{U(Y),V_{k}(Y)\}$ for all $k \in \calM$.
\cite{Hjort2003} showed that,
\begin{align}\label{eq:hjortasymp}\sqrt{n}(\sum_{k}w_{k}\hmu_{k} - \mu_{\true}) \toD \sum_{k}w_{k}\Lambda_{k},
\end{align}
where
\begin{align*} \Lambda_{k} = \begin{pmatrix}
\pa\mu(\bbeta_{0})/\pa \btheta \\ \pa\mu(\bbeta_{0})/\pa \bgamma_{k}
\end{pmatrix}^{\top} \left\{\matJ_{k}^{-1}\begin{pmatrix}
\matJ_{01}\bdelta \\ \pi_{k} \matJ_{11}\delta
\end{pmatrix} + \matJ_{k}^{-1}\begin{pmatrix}
\sqrt{n}(\Ubar_{n}- \bbE U_{k}(Y_{1}))\\ \sqrt{n}(\Vbar_{n,k} - \bbE V_{k}(Y_{1}))
\end{pmatrix}\right\} - \bigg(\dfrac{\pa\mu(\bbeta_{0})}{\pa \bgamma}\bigg)^{\top}\bdelta.
\end{align*}
Here, $\pi_{k}\in \Re^{|M_{k}|\times q}$ is the projection matrix that projects any vector $\bu \in \Re^{q}$ to $\bu_{k}\in \Re^{|M_{k}|}$ with indices as given by $M_{k} \in \calM$.

The following corollary states that the result in (\ref{eq:hjortasymp}) can be directly obtained from Theorem \ref{th:main2} and thus Theorem \ref{th:main2} covers the special setting (\ref{eq:mispec}) of \cite{Hjort2003}.  A proof of the corollary can be found in the Appendix. 
\begin{corollary}\label{cor:match}
Under the misspecification model (\ref{eq:mispec}), the asymptotic bias and variance in (\ref{eq:hjortasymp}) matches those in Theorem \ref{th:main2}.
\end{corollary}

\subsection{Selection of Weights in Frequentist Model Averaging}\label{sec:3}

Model averaging acknowledges the uncertainty caused by model selection and tackles the problem by weighting all models under consideration. To make it effective,
it is desirable that the weights can reflect the impact of each candidate model, which can be achieved by properly assigning a weight to each candidate model.
If model $k'$ is more likely to impact or is more plausible than the model $k$,  its associated weight $w_{k'}$ should be no smaller than $w_{k}$ for the model $k$. In our development, we propose to measure the strength of a model by its mean squared error, based on which we obtain a set of data-adaptive weights by minimizing the mean squared error of the combined model averaging estimator.
A similar scheme was developed in \cite{Liang2011}, where the authors minimized an unbiased estimator of mean squared error to obtain their optimal weights. However, their work was done for the linear models. As in \cite{Liang2011}, we assume that the true model is included in the set of candidate models in the development of our weighing scheme.

Recall Theorem \ref{th:main2},  the asymptotic mean squared error (AMSE) of $\bmu(\tbbetak)$  is, 
\begin{equation}\label{eq:AMSE} 
Q(\bw) = {\rm trace}\left((\sum_{k \in \calM} w_{k}\{\bmu(\tbbetakstar) - \bmu(\bbetatrue)\})^{\otimes 2}+ \dfrac{1}{n}\bSigma_{w}\right),
\end{equation}
for any given set of weights. 
However, this quantity depends on the unknown parameter $\bbetatrue$, so we instead  consider its estimate $\widehat{Q}_n(\bw)$. Assume that
we estimate $\bbetatrue$ consistently and the estimate is, say,  $\hbbeta_{cons}$. Then,  $Q(\bw)$ in (\ref{eq:AMSE}) can be consistently estimated by $\widehat{Q}_n(\bw) = Q(\bw) \big|_{\bbetatrue= \hbbeta_{cons}}$. We propose to obtain a set of data adaptive weights $\bw_{n}^{*}$ by minimizing $\widehat{Q}_{n}(\bw)$: 
$$
\bw_{n}^{*} = \argmin_{\bw} \, \widehat{Q}_{n}(\bw).
$$
The numerical performance of the proposed averaging estimators will be evaluated in Section 4. In the next section, we illustrate the procedure in the linear and logistic models in details.

\section{Model Averaging and Weight Selection in Regression  Models}
We now discuss the model averaging estimator described in Section \ref{sec:2} for generalized linear models (GLM).
Specifically, let $ \bbE y_{i}=g(\bx^{\top}_{i}\bbeta)$ where  $g$ is a given link function connecting the mean and the linear predictor $\eta_{i}=\bx^{\top}_{i}\bbeta$.
We consider a set $\calM$ of $2^{q}$ models. Suppose we want to estimate a function $\bmu(\bbeta)$ and, as defined in (\ref{fmae}), the final model averaging estimator is given by $\hbmu_{ave} = \sum_{k \in \calM} w_{k}\bmu(\tbbetak)$.
Since the set up for Theorem \ref{th:main2} is for a general parametric model, the same asymptotic convergence results hold for GLM models. In particular we verify condition (\textbf{A}1) and discuss the data-driven weight choices below in two special cases: linear regression and logistic models.

\subsection{Prediction in Linear Regression Models}

We first derive the model averaging estimator in the linear regression framework:
\[
   \by=\matX \bbeta +\bepsa,
\]
where $\matX\in \Re^{n\times (p+1)}$ is a non-random design matrix of full column rank; i.e., $\rank(\matX) = p+1$, and $\bepsa \sim \calN(\bzero,\sigma^2 \matI_{n})$.

Let $\calM= \{M_{k}\}^{|\calM|}_{k=1}$ be the set of candidate models. Here $M_{k}$ denotes a particular set of features having cardinality $|M_{k}|$. Define $\matX_{k}\in \Re^{n \times |M_{k}|}, 1\leq k \leq |\calM|$ as the design matrix of the $k^{\rm th}$ candidate model with the features in $M_{k}$. We consider
zero-augmentation of the parameter set $\bbetak$ for all $k$.
Let $\tmatX_{k}\in \Re^{n \times (p+1)}$ be the augmented version of $\matX_{k}$ with the missing columns replaced by the $\bzero$ vector. In our analysis, all the candidate models contain the intercept term corresponding to $\beta_{0}$. With the rest of the $p$ components, we can construct $2^{p}$ candidate models, all of which are included in our analysis.

Let us fix a $\bxstar\in \Re^{p+1}$. Define $\bx^{*}_{k}\in \Re^{|M_{k}|}$ so that $\bxkstar$ consists of those components of $\bxstar$ indexed by $M_{k}\in \calM$. Consider the particular choice of the function $\mu: \Re^{p+1} \rightarrow \Re^{}$ so that for $\bb\in \Re^{p+1}$, $\mu(\bb)=\bxstar^{\top}\bb$. Clearly the $\nabla\mu(\bbeta)= \bxstar$. For the following discussion, we are interested in the model averaging estimator of $\mu(\bbetatrue)$ = $\bxstar^{\top}\bbetatrue$, which is given by $\hmu_{ave} = \sum_{k}w_{k}\bxkstar^{\top}\hbbetak$ with $w_{k}\geq 0$ and $\sum_{k}w_{k}=1$. In the simulations, we will use $\bxstar$ generated from the known covariate distribution, while for the real data, we split the whole data set into a training set and a test set and $\bxstar$ will be set to be the covariate in the test~set.

For the $k^{\rm th}$ candidate model with $\bbetak\in \Re^{|M_{k}|}$, the score function is given by $\bell'_{k}(\bbetak) = \matX_{k}^{\top}(\by-\matX_{k}\bbetak)$ and $\matH_{k}$ is given by $\matH_{k}= -(1/n)\matX^{\top}_{k}\matX_{k}$; note that this follows from the definition immediately preceding condition (\textbf{A}1). 
Thus our Hessian matrix satisfies the condition as it does not depend on $\by$. Similarly we note that regarding condition (\textbf{A}1), 
\begin{equation}
|\nabla \mu(\tbbetakstar)\matH_{k}^{-1}\ell'_{k;i}(\bbetakstar)|
= \left|(y_{i}-[\matX_{k}]^{\top}_{i,\Cdot}\bbetakstar) \ \bxkstar^{\top}(\matX^{\top}_{k}\matX_{k}/n)^{-1}[\matX_{k}]_{i,\Cdot}\right|\\
 = |c_{ik}(\varepsilon_{i} + A_{ik})|, \nonumber
\end{equation}
where $c_{ik} = \bxkstar^{\top} (\matX^{\top}_{k}\matX_{k}/n)^{-1}[\matX_{k}]_{i,\Cdot}$ and $A_{ik} = \bx^{\top}_{i}\bbetatrue-[\matX_{k}]^{\top}_{i,\Cdot}\bbetakstar$ are fixed constants, and $[\matX_{k}]_{i,\Cdot}$ is the $i$th column of the matrix  $\matX_{k}^{\top}$.
Note that $\varepsilon_{i}\sim \calN(0,\sigma^{2})$. The condition (\textbf{A}1)   is satisfied if, for any arbitrary $\epsa>0$,
\[\lim _{n \rightarrow \infty}\dfrac{1}{n}\max_{1\leq i \leq  n} \bbE \left\{\max_{k\in \calM} |c_{ik}(\varepsilon_{i} + A_{ik})|\right\}^{2} \bbI\left\{ \max_{k\in \calM}|c_{ik}(\varepsilon_{i} + A_{ik})|> \sqrt{n}\epsa\right\} = 0.\]
Moreover, if $|c_{ik}|\leq C$ for some fixed constant $C>0$,  the condition is further reduced to,
\[\lim _{n \rightarrow \infty}\dfrac{1}{n}\max_{1\leq i \leq  n} \bbE \left\{\max_{k\in \calM} |\varepsilon_{i} + A_{ik}|\right\}^{2} \bbI\left\{ \max_{k\in \calM}|\varepsilon_{i} + A_{ik}|> \sqrt{n}\epsa\right\} = 0.\]
It is appropriate to note that we can have a bound of $c_{ik}$ as
\begin{align*}
\max_{k}|c_{ik}| = \max_{k}  |\bxkstar^{\top}(\matX^{\top}_{k}\matX_{k}/n)^{-1}[\matX_{k}]_{i,\Cdot}|
 \leq \norm{\bxstar}{}\norm{\bx_{i}}{}\max_{k} \dfrac{1}{\lambda_{min}(\matX^{\top}_{k}\matX_{k}/n)}.
\end{align*}
Here $\lambda_{min}(\matB)$ denotes the smallest singular value of matrix $\matB$. Now by an application of Cauchy-Schwarz inequality,
\begin{align}
& \dfrac{1}{n}\bbE \left\{\max_{k\in \calM} |\varepsilon_{i} + A_{ik}|\right\}^{2} \bbI\left\{ \max_{k\in \calM}|\varepsilon_{i} + A_{ik}|> \sqrt{n}\epsa\right\}\notag\\
& \qquad \leq \dfrac{1}{n} \left\{\bbE\max_{k\in \calM} |\varepsilon_{i} + A_{ik}|^{4} \right\}^{1/2} \left\{\ \bbP (\max_{k\in \calM} |\varepsilon_{i} + A_{ik}| > \sqrt{n}\epsa)\right\}^{1/2}\notag\\
& \qquad \leq \dfrac{1}{n}\sum_{k \in M} \bbE (\varepsilon_{i} + A_{ik})^{4} \left\{\sum_{k \in \calM} \bbP(|\varepsilon_{i} + A_{ik}|> \sqrt{n}\epsa)\right\}^{1/2}\notag\\
& \qquad \leq \left\{\dfrac{A^{4}_{ik}}{n^{2}} + 6\dfrac{A^{2}_{ik}\sigma^{2}}{n^{2}} + \dfrac{3\sigma^{4}}{n^{2}} \right\}^{1/2} \left\{\sum_{k \in \calM}\bbP(|\varepsilon_{i}|> \sqrt{n}\epsa - |A_{ik}|)\right\}^{1/2}\label{eq:fincond-lin}.
\end{align}
Thus it follows that for $|\calM|$ finite, as $n$ goes to infinity, the right hand side of  (\ref{eq:fincond-lin}) goes to zero and thus Condition  (\textbf{A}1) is satisfied.

The MLE of $\beta_k$ in the $k^{\rm th}$ model is given by $\hbbetak= (\matX^{\top}_{k}\matX_{k})^{-1}\matX^{\top}_{k}\by$. Let $\bbetakstar$ be such that $\bbE \bell_{k}'(\bbetakstar) = \bzero$; $\bbE \bell_{k}'(\bbeta_{k})$ being the score function of the $k^{\rm th}$ model, solving which we find that,
\begin{align}\label{eq:linreg-bbetakstar}
\bbetakstar = (\matX^{\top}_{k}\matX_{k})^{-1}\matX^{\top}_{k}\matX\bbetatrue.
\end{align}

\noindent As discussed in Section \ref{subsec:notation}, the entire set of candidate models can be divided into  two categories. The $1^{\rm st}$ category contains the ones that are biased and is denoted by $\calM_{\notin}$ and the second category contains ones that are not and is denoted by $\calM_{\in}$. So, for $k \in \calM_{\in}$ we have $\bbetakstar= \bbetatrue$,  whereas for $k \in \mathcal{M}_{\notin}$ we have $\bbetakstar \neq \bbetatrue$. Therefore the bias term of model averaging estimator $\hmu_{ave}$ can be written as,
\[
 \sum_{k\in \mathcal{M}_{\notin}} w_k (\bxkstar^{\top}\bbetakstar- \bxstar^{\top}\bbetatrue) =  \sum_{k\in \mathcal{M}_{\notin}} w_k\bxkstar^{\top}(\matX_k^{\top} \matX_k )^{-1} \matX_k^{\top} \matX\bbetatrue - \bxstar^{\top}\bbetatrue.
\]
Since the weights assigned to the models are unknown, we propose an estimate of the mean squared error (MSE) and minimize the MSE to obtain weights that would be assigned to the candidate models.  From Theorem \ref{th:main2}, the asymptotic mean squared error (MSE) of $\hmu_{ave}$ is given by
\begin{align}
  Q(\bw)
 &  =   \left[\left\{\sum_{k\in \calM_{\notin}}w_k(\bxkstar^{\top}\bbetakstar - \bxstar^{\top}\bbetatrue)\right\}^{2} \right.\notag \\
 & \qquad \qquad + \left. 	\dfrac{1}{n^{2}}\sum_{k \in \mathcal{M}} \sum_{k' \in \mathcal{M}}  \ w_k w_{k'}
   \bxkstar^{\top}\mathbf{H}_{k}^{-1}\ \bbE \bell'_{k}(\bbetatrue)\bell'_{k'}(\bbetatrue)^{\top}  \ {\mathbf{H}_{k'}^{-1}}^{\top} \bxstar_{k'}\right].\notag
	\label{eq:tracelinear}
\end{align}
Since $\mathbf{H}_{k}$ does not depend on $\by$, we focus on estimating  $\bbE \bell'_{k}(\bbetatrue)\bell'_{k'}(\bbetatrue)^{\top} $, which equals $
 \matX_{k}^{\top}   \bbE(\by-\matX\bbeta_{\true})(\by-\matX \bbeta_{\true})^{\top} \matX_{k'} =\sigma^{2} \matX^{\top}_{k}\matX_{k'}.$ 
It follows that
\begin{align*}
Q(\bw) &  =  \left\{  \sum_{k \in \calM_{\notin}}\sum_{k' \in \calM_{\notin}} w_{k}w_{k'}(\bxkstar^{\top}\bbetakstar - \bxstar^{\top}\bbetatrue)(\bxstar_{k'}^{\top}\bbeta^{*}_{k'} - \bxstar^{\top}\bbetatrue)\right.\notag \\
 & \qquad \qquad + \left. 	\sigma^{2}\sum_{k \in \mathcal{M}} \sum_{k' \in \mathcal{M}}  \ w_k w_{k'}
   \bxkstar^{\top}(\matX_{k}^{\top}\matX_{k})^{-1}\ \matX_{k}^{\top}\matX_{k'}  \ {(\matX_{k'}^{\top}\matX_{k'})^{-1}} \bxstar_{k'} \right\}.\notag
\end{align*}
Define the estimates of $\bbeta$ and $\sigma$ as
$\hbbeta_{full} = (\matX^{\top}\matX)^{-1}\matX^{\top}\by$ and $\hsigma_{full}^{2} = \norm{\by-\matX\hbbeta_{full}}{}^{2}/n$, respectively. 
Then $(\hbbeta_{full}, \hsigma_{full})$ are consistent estimates of $(\bbeta_{\true}, \sigma)$ under mild conditions. We therefore propose to estimate $Q(\bw)$ by 
\begin{eqnarray}
\widehat{Q}(\bw) & = & \sum_{k \in \calM}\sum_{k' \in \calM} w_{k}w_{k'}(\bxkstar^{\top}\hbbetak - \bxstar^{\top}\hbbeta_{full})(\bxstar_{k'}^{\top}\hbbeta_{k'} - \bxstar^{\top}\hbbeta_{full}) \nonumber \\
 && \quad \quad + \, \hsigma_{full}^{2}\sum_{k \in \mathcal{M}} \sum_{k' \in \mathcal{M}}  \ w_k w_{k'}
   \bxkstar^{\top}(\matX_{k}^{\top}\matX_{k})^{-1}\ \matX_{k}^{\top}\matX_{k'}  \ {(\matX_{k'}^{\top}\matX_{k'})^{-1}} \bxkstar. 
 \label{eq:1tracelin}
\end{eqnarray}
We obtain the weights for model averaging estimator $\bw= (w_1,\cdots,w_{|\calM|})$ such that $\widehat{Q}(\bw)$ in (\ref{eq:1tracelin}) is minimized.


\subsection{Estimation in Logistic Regression Framework}\label{subsec:logistic}

In this section we study the proposed model averaging estimation method under logistic regression models.
%
Let $\by \in \Re^{n}$ be $n$ independent copies of a dichotomous response variable $Y$ taking values 0/1. Let $\matX = (\bx_{1}, \cdots, \bx_{n})^{\top}\in \Re^{n \times (p+1)}$ be a set of features. The logit model is given by,
\[p_{i}= P(y_{i}=1|\matX)=\dfrac{\exp(\bx^{\top}_{i} \bbeta)}{1+\exp(\bx^{\top}_{i} \bbeta)}, \quad \forall i=1,\cdots, n, \]
where $\bbeta \in \Re^{p+1}$ are the set of unknown parameters of interest.
The log-likelihood for the logistic regression can be written as,
\begin{align}
\ell_{k}(\bbeta|\by,\matX)
 = & \log \prod_{i=1}^n  \dfrac{\exp(y_i \bx^{\top}_i \bbeta)}{1+\exp(\bx^{\top}_i \bbeta)} = \sum_{i=1}^n y_i\bx^{\top}_i \bbeta - \sum_{i=1}^n \log(1+\exp(\bx^{\top}_i \bbeta))\nonumber.
\end{align}
As before, let $\calM= \{M_{k}\}^{|\calM|}_{k=1}$ be the set of candidate models. Here $M_{k}$ denotes a particular set of features having cardinality $|M_{k}|$. Define $\matX_{k}
\in \Re^{n \times |M_{k}|}, 1\leq k \leq |\calM|$ as the design matrix of the $k^{\rm th}$ candidate model with the features in $M_{k}$. Denote by $[\matX_{k}]_{i,\Cdot}$
the $i$th column of the matrix  $\matX_{k}$, thus $[\matX_{k}]_{i,\Cdot}\in R^{|M_{k}|}$.
Let $\bbetak \in \Re^{|M_{k}|}$ be the parameter vector with components corresponding to the index set $M_{k}$. We consider zero-augmentation of the parameter set $\bbetak$ for all $k$ as was done for the linear regression models.

Again,  we consider estimation of a function of the form $p: \Re^{p+1} \rightarrow \Re$ given by
\begin{align} \label{eq:logit}
p(\bbeta) = \dfrac{\exp (\bxstar^{\top}{\bbeta})}{1 + \exp(\bxstar^{\top}\bbeta)}.
\end{align}
Let the unknown true parameter in our model be $\bbetatrue\in \Re^{p+1}$. Then $\bp_{\true}=\bp(\bbetatrue):= {\exp(\matX \bbeta_{\true})}/\{1+\exp(\matX \bbeta_{\true})\}\in \Re^{n}$ calculated component wise. To estimate the parameter $p_{\true}=p(\bbetatrue)$, we consider the model averaging estimator given by
\begin{align*}
\widehat p_{ave} = \sum_{k\in \calM} w_{k}p(\tbbetak),
\end{align*}
where $\tbbetak$ is the 0-augmented version of the MLE $\hbbetak$ of $\bbetak$ for the $k^{\rm th}$ model.
The score function for the $k^{\rm th}$ model is given by
\[
\bell^{'}_{k}(\bbeta_{k})=\sum_{i} y_i [\matX_{k}]_{i,\Cdot} - \sum_i \dfrac{\exp([\matX_{k}]_{i,\Cdot}^{\top} \bbetak)}{1+\exp([\matX_{k}]_{i,\Cdot}^{\top}\bbetak)}[\matX_{k}]_{i,\Cdot}= \matX_{k}^{\top} (\by-\bp_{k})\quad \forall\ 1\leq k \leq |\calM|,
\]
where $\bp_{k}= {\exp(\matX_{k}\bbetak)}/\{1+\exp(\matX_{k} \bbetak)\} \in \Re^{n}$.
The second derivative of the log-likelihood is given by
\[
\bell^{''}_{k}(\bbeta_{k}) = \sum_{i=1}^n \dfrac{\exp([\matX_{k}]_{i,\Cdot}^{\top} \bbetak)}{\{1+\exp([\matX_{k}]_{i,\Cdot}^{\top} \bbetak)\}^2}[\matX_{k}]_{i,\Cdot}[\matX_{k}]_{i,\Cdot}^{\top}= \matX_{k}^{\top} \matW_{k}(\matI_{n}-\matW_{k}) \matX_{k}\quad \forall\ 1\leq k \leq |\calM| ,
\]
where the weight matrix $\matW_{k} \in \Re^{n \times n}$ is a diagonal matrix defined as $\matW_{k} = \diag \big(p_{k;1}, \cdots, $ $p_{k;n}\big)$ with $p_{k;i} =  {\exp([\matX_{k}]_{i,\Cdot}^{\top} \bbetak)}\big/{\{1+\exp([\matX_{k}]_{i,\Cdot}^{\top} \bbetak)\}^2}$, for $i = 1, \ldots, n$.
Since  $\bell^{''}_{k}(\bbeta_{k})$ does not depend on $\by$, we have $\matH_{k} = (1/n)\bell^{''}_{k}(\bbeta_{k})$, for $1\leq k \leq |\calM|$. By simple algebra, it can be verified that Condition (\textbf{A}1) is satisfied for logistic regression model too.

To estimate the bias of the model averaging estimator, we define $\bbetakstar$ as the solution of the equation $\bbE[\bell'_{k}(\bbetak)]= \bbE\{\matX_k^{\top}(\by-\bp_{k})\}= \bzero$. That is, $\bbetakstar$ is a solution of
\begin{align}\label{eq:logitscore-k}
\matX_k^{\top} (\bp_{\true}-\bp_{k})= 0.
\end{align}
Denote by $\bp^{*}_{k}= {\exp(\matX_{k}\bbetakstar)}/\{1+\exp(\matX_{k} \bbetakstar)\} \in \Re^{n}$ calculated component wise. We have $\matX_k^{\top} (\bp_{\true}-\bp_{k}^{*})= 0$, and it follows that
\begin{align*}
\bbE \bell'_{k}(\bbetakstar)\bell'_{k'}(\bbetakstar)^{\top}
& = \matX_{k}^{\top}\bbE(\by-\bp^{*}_{k})(\by-\bp^{*}_{k'})^{\top}      \matX_{k'}\\
& = \matX_{k}^{\top}   \bbE\{(\by-\bp_{\true})- (\bp^{*}_{k}- \bp_{\true})\}\{(\by-\bp_{\true}) - (\bp^{*}_{k'}- \bp_{\true})\}^{\top}     \matX_{k'} \\
& = \matX_{k}^{\top}   \bbE (\by-\bp_{\true})(\by-\bp_{\true})^{\top} \matX_{k'} = \matX_{k}^{\top}  \matW^{\true} \matX_{k'}
\end{align*}
where $\matW^{\true}=var( \by-\bp_{\true}) = \bbE (\by-\bp_{\true})(\by-\bp_{\true})^{\top}$.
In addition, write $\matW^{*}_{k} = \diag(\bp^{*}_{k}) \in \Re^{n\times n}$. The gradient $\nabla p$ is given by
$
\nabla p(\bbetakstar) = p_k^*(1-p_k^*)\bxkstar,$ $1\leq k\leq |\calM|.$
 Thus, the MSE estimate is 
\begin{align*}
 Q(\bw)   & = \sum_{k \in \calM}\sum_{k' \in \calM} w_{k}w_{k'}(p^{*}_{k}-p_{\true})(p^{*}_{k'}-p_{\true})\notag \\
 & \qquad + \sum_{k \in \mathcal{M}} \sum_{k' \in \mathcal{M}}  \ w_{k} w_{k'}  p_k^*(1-p_k^*)\bxkstar^{\top}   \{\matX^{\top}_{k}\matW^{*}_{k}(\matI_{n}-\matW^{*}_{k})\matX_{k}\}^{-1}\\
 & \qquad \qquad 	 \times\  \matX_{k}^{\top} \matW^{\true}   \matX_{k'}\{\matX^{\top}_{k'}\matW^{*}_{k'}(\matI_{n}-\matW^{*}_{k'})\matX_{k'}\}^{-1} \bxkstar p_k^*(1-p_k^*)
  .\notag
\end{align*} 

However,  $Q(\bw)$ involves unknown $\bbetatrue$ and $\bbeta_k^{*}$. As in the linear regression model case, we use the full model to estimate $\bbeta_{\true}$ and denote by the estimator $\hbbeta_{full}$. Then, compute $\hbp_{full} = \bp(\hbbeta_{full})$ and $ \widehat p_{full} = p(\hbbeta_{full})$. The estimators $\hbp_k^{*}={\exp(\matX_{k}\hbbetak)}/\{1+\exp(\matX_{k} \hbbetak)\}    $ and $\widehat p_k^{*}={\exp(\bxkstar^{\top}\hbbetak)}/\{1+\exp(\bxkstar^{\top} \hbbetak)\}    $ are obtained by solving the equation
\begin{align}\label{eq:logitscore-k2}
\matX_k^{\top} (\hbp_{full}-\bp_{k})= 0,
\end{align}
using iterative re-weighted least squares (IRLS) method; cf., e.g., \cite{Holland2007}. Specifically, let ${\bbetak}^{(s)}$ be the solution of (\ref{eq:logitscore-k2}) at the $s^{\rm th}$ stage of the IRLS algorithm. The coefficients for the $(s+1)^{\rm th}$ stage is then given by
\begin{align*}
 {\bbetak}^{(s+1)} & = \left.{\bbeta}_k^{(s)} + \{\matX_k^{\top} \matW_{k}(\matI_{n}-\matW_{k}) \matX_k\}^{-1}\matX_k^{\top} \left\{ \dfrac{\exp(\matX \hbbeta_{full})}{1+\exp(\matX \hbbeta_{full})}-\dfrac{\exp(\matX_k {\bbetak}^{})}{1+\exp(\matX_k {\bbetak}^{})} \right\}\nonumber\right|_{\bbetak={\bbetak}^{(s)}},
\end{align*}
for $s = 0, 1, 2, ...$ When the algorithm converges, we obtain the estimate $\hbbetak$.  Putting together,
we estimate $Q(\bw)$ by 
\begin{align}\label{eq:logis_Q}
\hat Q(\bw)   & =\sum_{k \in \calM}\sum_{k' \in \calM} w_{k}w_{k'}(\widehat p^{*}_{k}-\widehat p_{full})(\widehat p^{*}_{k'}-\widehat p_{full})\notag \\
 & \qquad + \sum_{k \in \mathcal{M}} \sum_{k' \in \mathcal{M}}  \ w_{k} w_{k'} \widehat p_k^*(1-\widehat p_k^*)\bxkstar^{\top} \{\matX^{\top}_{k}\matW^{*}_{k}(\matI_{n}-\matW^{*}_{k})\matX_{k}\}^{-1}\notag\\
 & \qquad 	 \times\  \matX_{k}^{\top} \matW^{\true}   \matX_{k'}\{\matX^{\top}_{k'}\matW^{*}_{k'}(\matI_{n}-\matW^{*}_{k'})\matX_{k'}\}^{-1}\bxkstar \widehat p_k^*(1-\widehat p_k^*) \bigg |_{\bp_k^*= \hbp_k^*; \bp_{k'}^*= \hbp_{k'}^*; \bp_{\true} = \hbp_{full}}
   \biggr].
\end{align}



We can obtain $w_1,\cdots,w_N$ such that the estimated MSE $\hat Q(\bw)$ is minimized, similar to the development done in linear regression setup. These weights can be assigned to individual models for developing the model averaging estimator.

\section{Simulation Study \& Real Data Analysis}\label{sec:4}
\label{sec:simudata}
\subsection{Simulation Study I: bias and variance tradeoff}
\label{sec:simu1}

We study both finite and large sample behavior of the model averaging estimator under a regression setup: $\by = \matX\bbeta + \bepsa$ where $\by, \bepsa \in \Re^{n}$ and $\bbeta\in \Re^{p + 1}$. In the study, $p = 9$ and $\bbeta = (\beta_{0}, \beta_{1}, \cdots, \beta_{9})^{\mathsf{T}}$ where $\beta_{0}$ is the intercept coefficient. We assume that 5 parameters $(\beta_{0}, \cdots, \beta_{4})^{\mathsf{T}}$ are always included in all candidate models and the remaining parameters $(\beta_{5}, \cdots, \beta_{9})^{\mathsf{T}}$ may or may not be in a candidate model. For simulation of $\by$, first we set the true parameter (henceforth, referred to as $\bbeta^{*}$) as follows:
\[\bbeta^{*} = \underbrace{0.3,  0.3,  0.5,   0.1,  0.5}_{\text{Always Included}},  \underbrace{0.0, 0.6, 0.0, 0.1, 0.0}_{\text{Candidate Parameters}}.\]
For the design matrix, the first column of $\matX$ is chosen to be 1 (for interecept) and the rest are simulated independently from $\mathsf{N}(0,1)$ random variable. The final response $\by$ is obtained by adding independent Gaussian error $\epsa_{i} \sim \mathsf{N}(0,1)$ to each row. We also simulate $\bxstar = (1, \xstar_{1}, \cdots, \xstar_{9})^{\mathsf{T}}$ so that each element $\xstar_{j}$ is simulated from $\mathsf{N}(0, 1)$ and define our parameter of interest $\mu^{*} = {\bxstar}^{\mathsf{T}}\bbeta^{*}$.

\begin{figure}[t]
	\centering
	\textbf{Case A: True model among candidates}\\
  \includegraphics[width=0.9\textwidth]{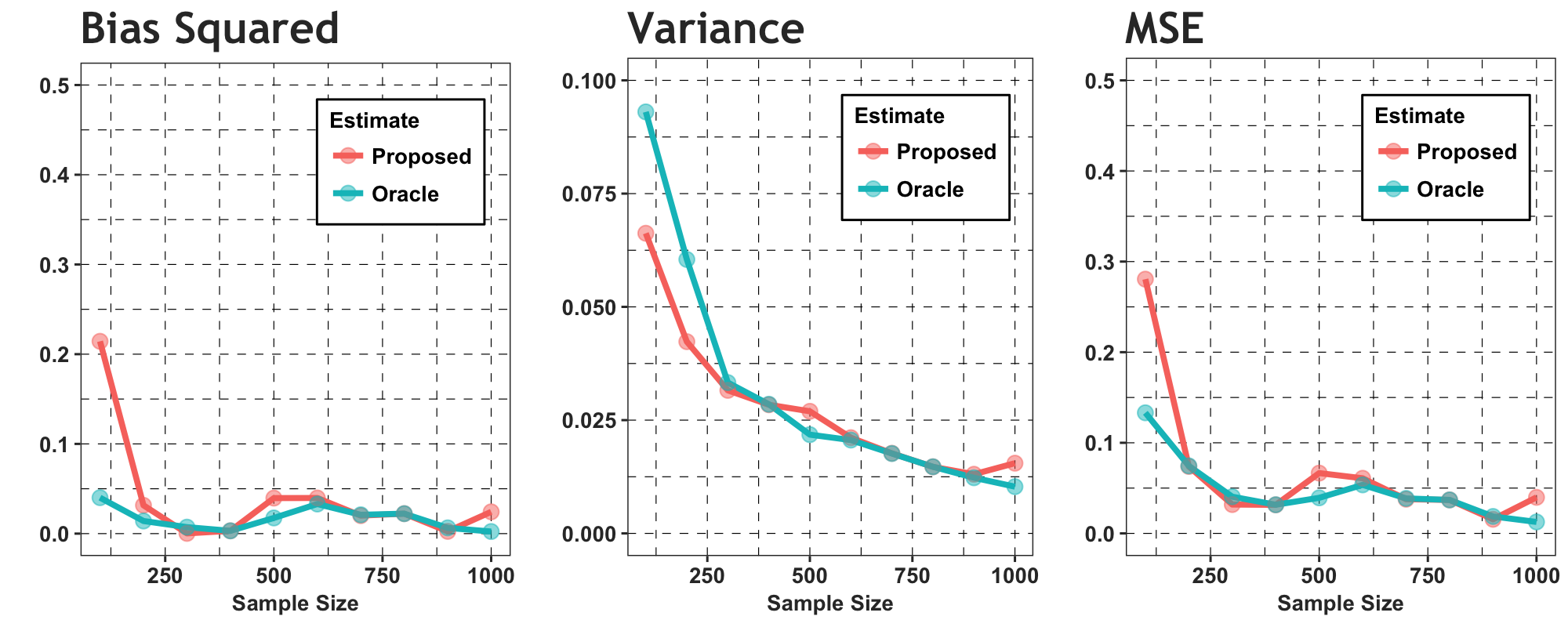}\\
  \textbf{Case B: True model not among candidates}\\
  \includegraphics[width=0.9\textwidth]{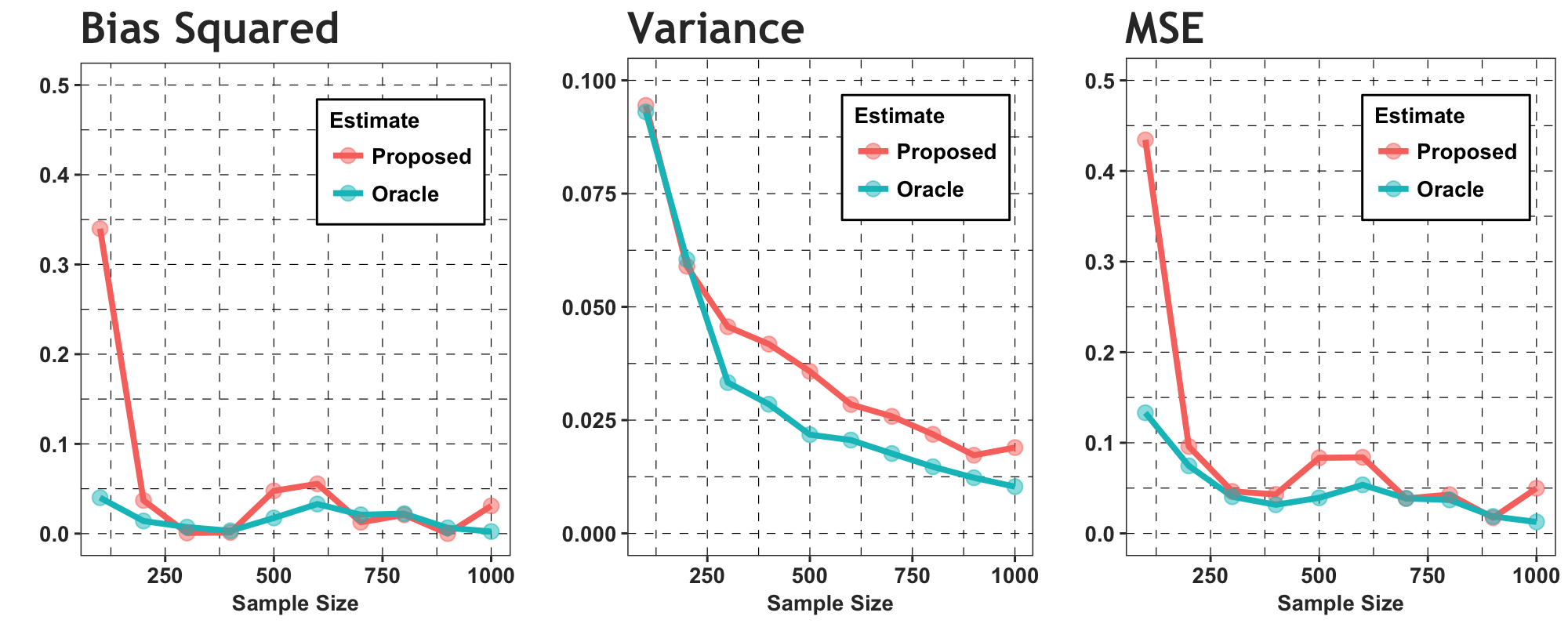}
  	\caption{Bias and  variance movement for the proposed model averaging and the oracle estimator of $\mu^{*}$. The true model is a sub-model of (nested within) some of the candidate models, but not included in the candidate model set in Case B.}
	\label{fig:pic1}
	\end{figure}
Clearly, based on all possible choices of last 5 parameters - there are a total of $2^{5} = 32$ candidate models. For ease of calculations we will consider the following 6 nested set of candidate models and the true/oracle model (represented pictorially):
\begin{align}
\label{eq:candiatemodels}
\kbordermatrix{
    & \mathbold{\beta_{5}} & \mathbold{\beta_{6}} & \mathbold{\beta_{7}} & \mathbold{\beta_{8}} & \mathbold{\beta_{9}} \\
    \text{\textbf{Candiate} 1} & \checkmark & \checkmark & \checkmark & \checkmark & \checkmark \\[-0.5em]
    \text{\textbf{Candiate} 2}  & \xmark & \checkmark & \checkmark & \checkmark & \checkmark \\[-0.5em]
    \text{\textbf{Candiate} 3}  & \xmark  & \xmark  & \checkmark & \checkmark & \checkmark \\[-0.5em]
    \text{\textbf{Candiate} 4}  & \xmark  & \xmark  & \xmark  & \checkmark & \checkmark \\[-0.5em]
    \text{\textbf{Candiate} 5}  & \xmark  & \xmark  & \xmark  & \xmark  & \checkmark
  \\[-0.5em]
  \text{\textbf{Candiate} 6}  & \xmark  & \xmark  & \xmark  & \xmark  & \xmark
  \\[-0.5em]
  \text{\textbf{Oracle } }  & \xmark  & \checkmark & \xmark  & \checkmark & \xmark
  }.
\end{align}
Note that the true model is a sub-model of candidate models 1 and 2, and candidate model 6 only contains the first 5 fixed parameters and none of the candidate parameters are included. We will consider two cases --- In \textbf{Case A} we will consider all 7 models in (\ref{eq:candiatemodels}) comprising of the 6 nested models and the true model; In \textbf{Case B}, we will only consider the first 6 nested models. We will compare our results with that of the \textit{oracle estimate}, where we know before-hand which parameters are non-zero and use a least -squared method to estimate $\bbeta$ and consequently $\mu^{*}$. We vary the sample size $n$ from 100 to 1000 and compare the bias and variance between the proposed and the oracle method.

In Figure \ref{fig:pic1}, we consider two cases: \textbf{Case A}, where the true (or oracle)-model is one of the candidate models and \textbf{Case B}, the true model is not one of the candidate models. In Figure \ref{fig:pic1} we compare the squared bias, variance and mean squared error movements as sample size is increased. In the top panel for \textbf{Case A}, the model-average estimator has less variance than the oracle estimate even for very small sample sizes which is to be expected; the reason being that the candidate model set contains the oracle as one of its candidates and further averaging reduces variances. In the bottom panel for \textbf{Case B}, with the increase in sample size, the variance of the proposed estimator decreases but is slightly higher compared to oracle. In both cases, the bias matches the oracle very closely as the sample size increases. It is suggestive from the plots in Figure \ref{fig:pic1} that in the linear regression setup, even when the candidate models do not include the true set of parameters, model averaging approaches the performance of oracle estimator in terms of bias and variance. We want to stress that this close performance of the model averaging estimator as compared to oracle is specific to this simple linear regression setup where the true model is a sub-model of some of the candidate models. In general, the question of \textit{whether the performance of model averaging estimator is close to the oracle},  would require separate investigation specific to the model and data at hand.



\subsection{Simulation Study II: Comparison with existing model averaging methods}

In this subsection we use both linear and logistic regression models to perform simulation studies to compare the performance of the frequentist model averaging estimator with the proposed weights with two existing model averaging methods by \cite{Hjort2003} and \citet{Liang2011}. The method by \cite{Hjort2003} (which we refer to as the FMA method) and the method by \citet{Liang2011} (which we refer to as the OPT method) are two well-studied approaches and both are also close to ours. The FMA medthod combines estimators from different models with the assumption that the data are coming from a local misspecification framework so the candidate model used has to have a bias of $\calO({1}/{\sqrt n})$ or less. We do not have this restriction in our proposed method. 
The OPT method proposes an unbiased estimate of MSE of the model averaging estimator and then the model averaging weights are obtained by minimizing the trace of the MSE estimate. The weight selection for OPT has been shown to exhibit optimality properties in terms of minimizing the MSE. However, their development is limited only to linear regression setting. 

\vskip5pt\noindent\textbf{Linear Regression:} In the linear regression setup, we work with a design similar to the one we described in Subsetion \ref{sec:simu1}. In particular, in the setup $\by = \matX\bbeta + \bepsa$ where $\by, \bepsa \in \Re^{n}$ and $\bbeta\in \Re^{p}$, we take $p = 4$ and $n = 100$; we denote $\bbeta = (\beta_{0}, \beta_{1}, \beta_{2}, \beta_{3})$ with $\beta_{0}$ being the coefficient for the intercept. In this setup the fixed parameter is $\beta_{0}$ (i.e. $k$ = 1) and the rest may or may not appear in the model (i.e. $m$ = 3). As before, we use $\bbeta^{*}$ to denote the true parameter. The elements of the design matrix $\matX$ is simulated independently from a $\sfN(0,1)$ distribution and the elements of the error vector $\bepsa$ is simulated independently as $\sfN(0, 1)$. 
\begin{table}[t]
	\centering
			\scalebox{0.8}{\begin{tabular}{cccclcclcclcc}
			\toprule
			\multicolumn{13}{c}{\textbf{Case A : True model among candidates}}\\[0.5em]
			\toprule
			\multirow{3}{*}{$\beta^{*}_{3}$} & \multirow{3}{*}{$\mu^{*}$} &\multicolumn{2}{c}{\textbf{(a) Proposed}} & &\multicolumn{2}{c}{\textbf{(b) OPT}} & & \multicolumn{2}{c}{\textbf{(c) FMA}} & &\multicolumn{2}{c}{\textbf{(d) Oracle}}\\
			 \cline{3-4}\cline{6-7}\cline{9-10}\cline{12-13}\\[-0.7em]
			 & &\textbf{Estimate} & \textbf{\textit{Error}} & &\textbf{Estimate} & \textbf{\textit{Error}} & &\textbf{Estimate} & \textbf{\textit{Error}} & &\textbf{Estimate} & \textbf{\textit{Error}}\\
			\midrule	
		0.001 & -0.192 & -0.059 & \textit{0.249} & &  -0.028 & \textit{0.231} & & 0.186 & \textit{0.400} & & -0.14 & \textit{0.221}\\
		0.005 & -0.196 & -0.062 & \textit{0.250} & & -0.032 & \textit{0.231} & & 0.184 & \textit{0.402} & & -0.145 & \textit{0.221}\\
		0.01 & -0.202 & -0.064 & \textit{0.252} & & -0.037 & \textit{0.232} & & 0.182 & \textit{0.404} & & -0.15 & \textit{0.221}\\
		0.05 & -0.243 & -0.103 & \textit{0.261} & & -0.075 & \textit{0.238} & & 0.165 & \textit{0.421} & & -0.192 & \textit{0.221}\\
		0.1 & -0.296 & -0.149 & \textit{0.268} & & -0.119 & \textit{0.248} & & 0.148 & \textit{0.445} & & -0.244 & \textit{0.221}\\
		0.5 & -0.714 & -0.599 & \textit{0.248} & & 0.104 & \textit{0.832} & & 0.129 & \textit{0.849} & & -0.662 & \textit{0.221}\\
		\toprule
			\multicolumn{13}{c}{\textbf{Case B: True model not among candidates}}\\[0.5em]
			\toprule
			\multirow{3}{*}{$\beta^{*}_{3}$} & \multirow{3}{*}{$\mu^{*}$} &\multicolumn{2}{c}{\textbf{(a) Proposed}} & &\multicolumn{2}{c}{\textbf{(b) OPT}} & & \multicolumn{2}{c}{\textbf{(c) FMA}} & &\multicolumn{2}{c}{\textbf{(d) Oracle}}\\
			 \cline{3-4}\cline{6-7}\cline{9-10}\cline{12-13}\\[-0.7em]
			 & &\textbf{Estimate} & \textbf{\textit{Error}} & &\textbf{Estimate} & \textbf{\textit{Error}} & &\textbf{Estimate} & \textbf{\textit{Error}} & &\textbf{Estimate} & \textbf{\textit{Error}}\\
			\midrule	
		0.001 & -0.192 & -0.058 & \textit{0.246} & & 0.063 & \textit{0.373} & & 0.217 & \textit{0.430} & & -0.14 & \textit{0.221}\\
		0.005 & -0.196 & -0.06 & \textit{0.248} & & 0.063 & \textit{0.375} & & 0.217 & \textit{0.433} & & -0.145 & \textit{0.221}\\
		0.01 & -0.202 & -0.061 & \textit{0.249} & &  0.063 & \textit{0.379} & & 0.216 & \textit{0.438} & & -0.15 & \textit{0.221}\\
		0.05 & -0.243 & -0.091 & \textit{0.259} & & 0.064 & \textit{0.406} & & 0.21 & \textit{0.472} & & -0.192 & \textit{0.221}\\
		0.1 & -0.296 & -0.112 & \textit{0.276} & & 0.066 & \textit{0.444} &  & 0.202 & \textit{0.515} & & -0.244 & \textit{0.221}\\
		0.5 & -0.714 & -0.073 & \textit{0.663} & & 0.077 & \textit{0.817} & & 0.129 & \textit{0.849} & & -0.662 & \textit{0.221}\\
		\hline
		\bottomrule
		\end{tabular}}
			\caption{(\textbf{Linear Regression}) Mean squared error for estimation of $\mu^{*}$ for the (a) model averaging estimator with proposed weights, (b) model averaging estimator with Liang's (\cite{Liang2011}) weights, (c) Hjort's (\cite{Hjort2003}) model averaging estimator with AIC based weights, and (d) oracle estimator. Here, in the top table, the candidate models include the true set of parameters (\textbf{Case A}) and in the bottom table true set of parameters is not included (\textbf{Case B}) - as described in (\ref{eq:candimods}). }
\label{tab:mse_oracInNotIn}
\end{table}

In this simulation setup, the estimand of interest is the following:
\begin{align*}
\mu^{*} = \bxstar^{\T}\bbeta^{*}, \text{ where } \bxstar \sim \sfN_{p}(\bzero, \matI_{4}).
\end{align*} 
For our specific example, we have $\bxstar = (1, -1.855445,  -1.018565,  -1.045111)$ and the true parameter $\bbeta^{*} = (0.3, 0.1, 0.3, \beta^{*}_{3})$. In the following we will vary the value of $\beta^{*}_{3}$ in the set $\{0.001, 0.005, 0.01, 0.05, 0.1, 0.5\}$ and compare the performances of different methods. As before, we will consider two different sets of candidate models; 
\begin{align}\label{eq:candimods}
\begin{array}{ccl}
\textbf{Case A } & : &\{\beta_{0}\},\ \{\beta_{0}, \beta_{1}\},\ \{\beta_{0}, \beta_{1}, \beta_{2}\},\ \{\beta_{0}, \beta_{1}, \beta_{2}, \beta_{3}\}\\
\textbf{Case B } & : &\{\beta_{0}\},\ \{\beta_{0}, \beta_{1}\},\ \{\beta_{0}, \beta_{1}, \beta_{2}\}.
\end{array}
\end{align}
Note that in \textbf{Case A}, the true parameter set is included in the model while in \textbf{Case B}, the true parameter set is not included. In fact, \textbf{Case B} represents a typical scenario where researchers are not even aware of the presence of the existence of the feature corresponding to $\beta_{3}$ and hence is working under a mis-specified model.

In Table \ref{tab:mse_oracInNotIn} the performances of different methods are compared for \textbf{Case A} (at the top) and \textbf{Case B} (bottom). For each separate choice of $\beta^{*}_{3}$, we performed 10 simulations and reported their averages in Table \ref{tab:mse_oracInNotIn} along with the root mean square error (in italics). Specifically the error for this simulation setup was defined as,
\[
\text{Error}  \ = \ \sqrt{({1}/{10})\sum_{k = 1}^{10}|\hmu_{k}-\mu^{*}|^{2}},
\]
whre $\hmu_{k}$ is the estimate corresponding to a specifc method at the $k^{\rm th}$ simulation. In Table \ref{tab:mse_oracInNotIn} \textbf{Case A} we compare the metrhods when $\beta_{3}$ is included in the largest candiate model while in \textbf{Case B},  $\beta_{3}$ is not considered in any of the candidate models. From Table \ref{tab:mse_oracInNotIn} \textbf{Case A}, it can be seen that in the finite sample framework ($n = 100$), the performances of proposed model-average estimator and OPT are similar and both outperform FMA. Moreover with the increase in magnitude of $\beta^{*}_{3}$ to 0.5, proposed model averaging method outperforms both FMA and OPT.  On the other hand, the setup in Table \ref{tab:mse_oracInNotIn} \textbf{Case B} 
shows that with the increase in $\beta^{*}_{3}$, the estimation error increases consistently for all three methods. Nevertheless, our proposed method clearly outperforms the competing methods in this scenario for all $\beta^{*}_{3}$ values. We also remark that the proposed method performs well up till $\beta_{3} = 0.1$, but the error jumps for the larger signal with $\beta_{3} = 0.5$. This is expected since $\beta_{3}$ is not considered in any of the candidate models and the extent of model mis-specification is large at $\beta^{*}_{3} = 0.5$. 

\vskip5pt\noindent\textbf{Logistic Regression:} 
\begin{table}[t]
	\centering
			\scalebox{0.8}{\begin{tabular}{cccclcclcc}
			\toprule
			\multicolumn{10}{c}{\textbf{Case A: True model among candidates}}\\[0.5em]
			\toprule
			\multirow{3}{*}{$\beta^{*}_{3}$} & \multirow{3}{*}{$p^{*}$} &\multicolumn{2}{c}{\textbf{(a) Proposed}} & &\multicolumn{2}{c}{\textbf{(b) FMA}} & & \multicolumn{2}{c}{\textbf{(c) Oracle}}\\
			 \cline{3-4}\cline{6-7}\cline{9-10}\\[-0.7em]
			 & &\textbf{Estimate} & \textbf{\textit{Error}} & &\textbf{Estimate} & \textbf{\textit{Error}} & &\textbf{Estimate} & \textbf{\textit{Error}} \\
			\midrule		
			0.001 & 0.452 & 0.457 & \textit{0.102} & & 0.515 & \textit{0.115} & & 0.418 & \textit{0.114}\\
			0.005 & 0.451 & 0.457 & \textit{0.102} & & 0.515 & \textit{0.116} & & 0.418 & \textit{0.114}\\
			0.01 & 0.45 & 0.46 & \textit{0.101} & & 0.518 & \textit{0.116} & & 0.419 & \textit{0.110}\\
			0.05 & 0.439 & 0.46 & \textit{0.126} & & 0.529 & \textit{0.126} & & 0.428 & \textit{0.139}\\
			0.1 & 0.427 & 0.44 & \textit{0.147} & & 0.534 & \textit{0.135} & & 0.398 & \textit{0.145}\\
			0.5 & 0.329 & 0.386 & \textit{0.173} & & 0.547 & \textit{0.230} & & 0.357 & \textit{0.166}\\
		\midrule
		\multicolumn{10}{c}{\textbf{Case B: True model not among candidates}}\\[0.5em]
			\toprule
			\multirow{3}{*}{$\beta^{*}_{3}$} & \multirow{3}{*}{$p^{*}$} &\multicolumn{2}{c}{\textbf{(a) Proposed}} & &\multicolumn{2}{c}{\textbf{(b) FMA}} & & \multicolumn{2}{c}{\textbf{(c) Oracle}}\\
			 \cline{3-4}\cline{6-7}\cline{9-10}\\[-0.7em]
			 & &\textbf{Estimate} & \textbf{\textit{Error}} & &\textbf{Estimate} & \textbf{\textit{Error}} & &\textbf{Estimate} & \textbf{\textit{Error}} \\
			\midrule		
			0.001 & 0.452 & 0.475 & \textit{0.093} & & 0.543 & \textit{0.123} & & 0.418 & \textit{0.114}\\
			0.005 & 0.451 & 0.475 & \textit{0.093} & & 0.543 & \textit{0.124} & & 0.418 & \textit{0.114}\\
			0.01 & 0.45 & 0.478 & \textit{0.092} & & 0.546 & \textit{0.125} & & 0.419 & \textit{0.110}\\
			0.05 & 0.439 & 0.473 & \textit{0.118} & & 0.554 & \textit{0.134} & & 0.428 & \textit{0.139}\\
			0.1 & 0.427 & 0.456 & \textit{0.135} & & 0.561 & \textit{0.152} & & 0.398 & \textit{0.145}\\
			0.5 & 0.329 & 0.478 & \textit{0.187} & & 0.56 & \textit{0.239} & & 0.357 & \textit{0.166}\\
			\hline
		\bottomrule
		\end{tabular}}
			\caption{(\textbf{Logistic Regression}) Estimation of $p^{*}$ for the (a) model averaging estimator with proposed weights (b) Hjort's (\cite{Hjort2003}) model averaging estimator with AIC based weights, and (c) oracle estimator. Here, in the top table, the candidate models include the true set of parameters (\textbf{Case A}) and in the bottom table true set of parameters is not included (\textbf{Case B}) - as described in (\ref{eq:candimods}).}
\label{tab:err_oracInNotIn}
\end{table}
We now describe the efficacy of the proposed methodology for logistic regression setup and compare its performance with Hjort's FMA method (\cite{Hjort2003}). The logit model is given by,
\begin{align}
p_{i}= P(y_{i}=1|\matX)=\dfrac{\exp(\bx^{\top}_{i} \bbeta)}{1+\exp(\bx^{\top}_{i} \bbeta)}, \quad \forall i=1,\cdots, n,
\end{align}
where $\matX = [\bx_{1}, \cdots,\bx_{i}, \cdots, \bx_{n}]^{\T} \in \Re^{n\times p}$ where $\bx_{i}\in \Re^{p}$ and $\bbeta\in \Re^{p}$. We take $n=100$ and $p$ = 4 where the intercept is always included ($k=1$) and the rest of the parameters can be varied in forming candidate models ($m = 3$). As in the linear regression simulation setup, the elements of $\matX$ is simulated independently from $\sfN(0, 1)$ distribution. In this setup, the true value of the parameter $\bbeta$ is set as $\bbeta^{*} = (0.3, 0.1, 0.3, \beta_{3}^*)$, where we vary the value of $\beta^{*}_{3}$ (as before) in the set $\{0.001, 0.005, 0.01, 0.05, 0.1, 0.5\}$. For this logistic regression setup,  our estimand of interest is as follows:
\begin{align}\label{eq:estimand_logistic}
p^{*} = \exp(\eta^{*})/(1+\exp(\eta^{*})) \text{ where } \eta^{*} = \bxstar^{\T}\bbeta^{*} \text{ and } \bxstar \sim \sfN_{4}(\bzero, \matI_{4}).
\end{align}

As in the regression setup, we set $\bxstar = [1.0, -1.86,  -1.019,  -1.045]$. Note that the specifics of model averaging estimator for the estimand in (\ref{eq:estimand_logistic}) has been described in detail in Section \ref{subsec:logistic}. Specifically, (\ref{eq:logis_Q}) describes the MSE function to be minimized for optimal weights. We compare our prposed method with Hjort's FMA method (\cite{Hjort2003}) and the oracle estimate. As in the linear regression setup, we consider two cases namely, \textbf{Case A} and \textbf{Case B}; see (\ref{eq:candimods}) for more details. The results for both \textbf{Case A} and \textbf{Case B} are summarized in Table \ref{tab:err_oracInNotIn}. We define the error metric as, 
\[
\text{Error}  \ = \ \sqrt{({1}/{10})\sum_{k = 1}^{10}|\phat_{k}-p^{*}|^{2}},
\]
whre $\phat_{k}$ is the estimate corresponding to a specifc method at the $k^{\rm th}$ simulation.
As in the linear regression setup, for the logistic regression as well, we see that the proposed method performs better than Hjort's method using AIC-based weights in both cases across all $\beta^{*}_{3}$ values. For \textbf{Case A}, the performance of our proposed method matches that of the oracle and the differences are with the margin of error.  For \textbf{Case B}, the performance of our proposed method tracks well with the oracle until the signal strength of $\beta^{*}_{3}$ is increased to 0.5, in which case the estimation error increases.


\subsection{Analysis of Prostate Cancer Data.}

The data for this example come from a study by \cite{Stamey1989}. They examined the relationship between the level of prostate-specific antigen and
a number of clinical measures in men who were about to receive a radical prostatectomy. As a regression problem, the response variable is \textsf{lpsa}, the level of prostate-specific antigen, with values ranging from -0.43 to 5.58. The predictor variables (clinical measures) are log cancer volume (\textsf{lcavol}), log prostate weight (\textsf{lweight}),
\textsf{age}, log of the amount of benign prostatic hyperplasia (\textsf{lbph}), seminal vesicle invasion (\textsf{svi}), log of capsular penetration (\textsf{lcp}),Gleason score (\textsf{gleason}), and percent of Gleason scores 4 or 5 (\textsf{pgg45}). Here \textsf{svi} is a binary variable, and \textsf{gleason} is an ordered categorical variable.

We considered a best-subset model selection approach using an all-subsets search. 
In this model selection approach, the estimated prediction error is obtained using a crude cross-validation method:  
the dataset is divided randomly into a training set of size 67 and a test set of size 30. The training set is used to select a model and then the test set is used to compute the  prediction error, averaging over all 30 points. We repeat the process five times and average over the five prediction errors.

\begin{table*}[h]
	\centering
		\begin{tabular} {lcc}
		\toprule
\textbf{Method Used} & \textbf{Test Error}\\
\midrule

Model Selection (Best Subset Regression) & 0.487 \\

Model Averaging (Proposed Weights) &  0.453  \\

 Model Averaging (AIC Weights) &  0.987  \\

Full Model & 1.272 \\

\bottomrule
			
		\end{tabular}
\caption{Prediction Error for different methods for prostate cancer data.}
	\label{tab:PredictionErrorForDifferentMethods}
\end{table*}

We also considered the model averaging method using two different sets of weights: the proposed weights and also AIC-based weights. 
Using the proposed weights, the proposed approach assigned the most weights to the model with features \textsf{lcavol}, \textsf{lweight}, \textsf{svi}, \textsf{pgg45}, \textsf{lcp}, \textsf{gleason} and \textsf{lbph} and the model with \textsf{lcavol}, \textsf{lweight}, \textsf{svi}, \textsf{pgg45}, \textsf{lcp}, \textsf{gleason}, \textsf{lbph} and \textsf{age}. The procedure with AIC-based weights gives more weight to a smaller model containing \textsf{lcavol} and \textsf{lweight}. We used the same crude cross-validation method as above, with a training set of size 67 and a test set of size 30. The training set is used to obtain the model averaging estimates and then the test set is used to compute the prediction error, averaging over all 30 points. We repeat the process five times and average over the five prediction errors.

\begin{figure}[t]
	\centering
		\includegraphics[width=1\textwidth]{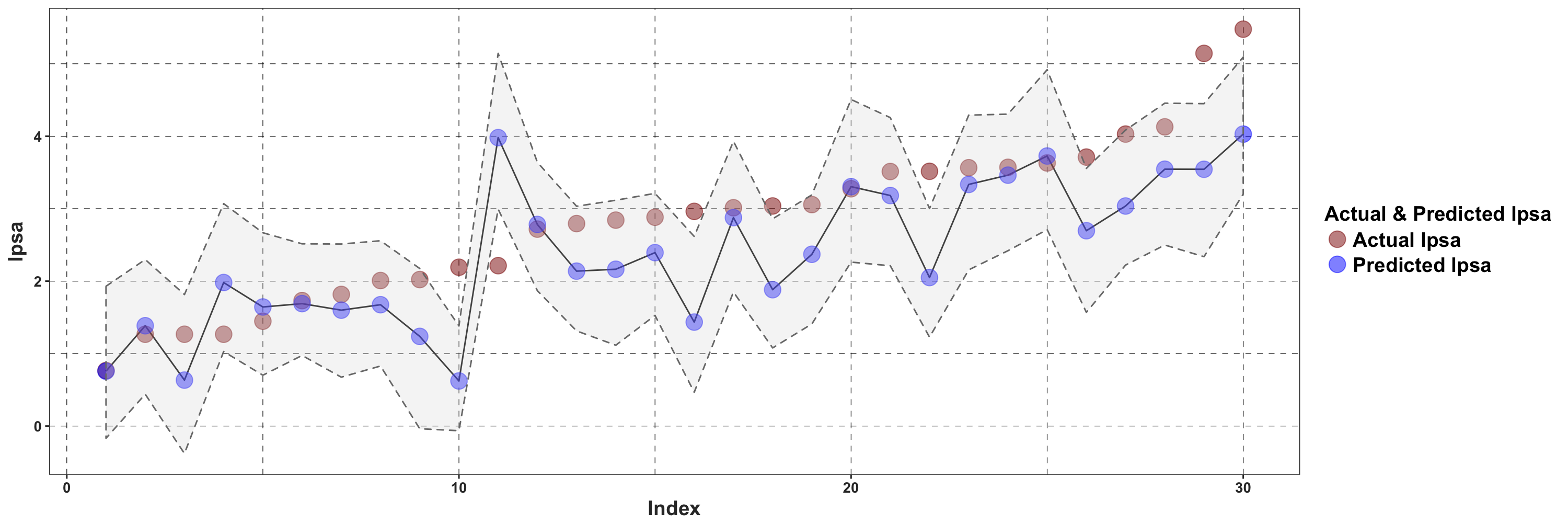}
		\caption{Actual and predicted level of \textsf{lpsa} (level of prostate-specific antigen) based on the prostate cancer data from \cite{Stamey1989}. In the x-axis the indices of the 30 observations are noted. In the y-axis we note the \textsf{lpsa} values. The red points indicate actual (observed) values while the blue points indicate predited values based on average of 50 replications. The gray band denotes the 90\% confidence interval.}
	\label{fig:pic3}
\end{figure}

Finally, as an illustration, we also plotted in Figure~\ref{fig:pic3}, a set of $90\%$ prediction intervals of antigen levels for one test dataset in one of our simulation runs. The x-axis is the index of the 30 observations in the test dataset. In order to get the prediction interval, we kept the test dataset fixed while in 50 different replications we selected a random subset of 50 observations from the training data (of original size 67) and applied the model averaging method to analyze the training data of size $50$ and use the result to predict the \textsf{lpsa} values for the test dataset. In order to construct the prediction interval we added to each predicted mean, a Gaussian noise with mean 0 and standard error equal to the estimated standard error from the full model denoted as $\hsigma_{full}$; see (\ref{eq:1tracelin}). In this  case the the estimate was $\hsigma_{full} = 0.599$. The upper and lower limits of the prediction band were calculated based on quantiles. As is clear from the plots, most of the observations fall within the 90\% prediction interval.

\section{Discussion}

In this paper, we propose a more general framework where the choice of true model is not fixed. The truth can be any one or a mixture of the candidate models. Models that have large biases are not excluded from our analysis. We study the behavior of frequentist model averaging estimator with an optimal weighting scheme to combine all the individual candidate models.
As an illustration, we derive the model averaging estimator in the linear and logistic regression framework. 
We also implement the weighting scheme proposed by \citet{Liang2011} and compare their performance to AIC based weights.
The simulation results indicate that under certain model specifications, the proposed estimator works better than \cite{Hjort2003}'s estimator. 

There are many ways a regression model can be misspecified. Misspecification in most cases is often interpreted as a case of left out variables or when the functional form of the model is not correctly specified.
In these instances, the normality assumption among random errors are violated.
This results in the estimates being biased as discussed in \cite{Giles1992}. These estimates can harm the decision making process, so one should be very attentive while
fitting and choosing models in the presence of misspecification. Many methods have been used to measure and limit misspecification in model fitting. Ramsey regression equation specification error test, discussed in \cite{jerry1977}, may help provide a test that is useful in a linear regression setup.

In model averaging, if the true model is not included in the set of candidate models, we end up using an estimate that is biased. If all the models are misspecified, the weights derived by AIC or by using a consistent or unbiased estimator of mean squared error are not optimal and should be with care. When the true model is not included in the analysis thus all the candidate models are wrong, there have been developments in model selection that takes care of the bias resulting from selection. See \cite{Hurvich1989, Hurvich1991}. A penalized version of AIC and BIC have been derived that performs better than other selection criteria. One can follow a similar path and derive the model averaging weights based on a sightly modified criteria.

Another problem with model averaging is that the number of optional parameters in analysis could be very high. For example, if there are $30$ parameters we could end up using as many as $2^{30}$candidate models. This may be time consuming and not ideal in certain fields of study. However, as suggested in this paper, a statistician can choose to use all or very few candidate models as per the scope of the study. This could be explored in further developments.

\renewcommand{\theequation}{{A.\arabic{equation}}}
\renewcommand{\thesection}{{\it \Alph{section}}}
\renewcommand{\thesubsection}{{\it A.\arabic{subsection}}}
\setcounter{equation}{0}
\setcounter{section}{0}
\setcounter{subsection}{0}

\section{Appendices}\label{sec:7}
\subsection*{A.1. Regularity Conditions and Assumptions}

In this section we state the regularity conditions that were used throughout the paper. We assume that the density function satisfies the following conditions.
\begin{enumerate}[(a)]
\item  $\Theta$ is an open subset of $\Re^{p}$, and the support of the density $f(y,\bbeta)$ is independent of $\bbeta$. 
\item  The true parameter value is an interior point of the parameter space.
\item  $\ellki'$ and $\telli''(\bbetakstar)$ exists and $\ellki'$ is a continuous function of $\bbeta$.
\item  $\bbE[\ellki']=0$ and $\bbE[\ellki' \ellki'^{\top}] = -\bbE[\telli''(\bbetakstar)]$. These conditions are standard conditions for asymptotic normality of maximum likelihood estimators.
\item  $ \lim_{n\rightarrow \infty} \dfrac{1}{n} \left[\tell''(\bbetakstar)\right] \rightarrow \mathbf{H}_{k}$ and $\mathbf{H}_{k}$ is positive definite.
\item  For some $\epsilon > 0 $, $\sum_i \bbE|\lambda'\ellki'(\bbetatrue)|^{2+\epsilon}/n^{(2+\epsilon)/2} \rightarrow 0$ for all $\epsilon \in \Re^{p}$.
\item  There exists $\epsilon > 0 $ and random variables $B_i(y_i)$,  $sup \left\{ |\telli''(\bbetakstar)|:|| t-\bbetatrue || \leq \epsilon  \right\} \leq B_i(y_i)$ and $\bbE|B_i(y_i)|^{1+\delta} \leq K$, where $\delta$ and $K$ are positive constants.
\end{enumerate}
We also assume that the variance matrix of the score statistic is finite and positive definite.

Consider a functional $\mu: \Re^{p + q} \rightarrow \Re$.  Define $\mu^{(\drop)}: \Re^{p+m} \rightarrow \Re$ as the same function as $\mu$ with only the $(q-m)$ corresponding arguments dropped.
For any $\bb = (b_{1}, \cdots, b_{p}, b_{p+1}, \cdots, b_{p +m})$ with $1\leq m \leq q$ define the \emph{$\bc$-augmented} version of $\bb$ as $\widetilde{\bb} = \{\bb, \bc\}\in \Re^{p+q}$ with some fixed $\bc\in \bar{\Re}^{q-m}$ inserted at the place of missing components. Let the indices of the missing components be $\{p +i_{1}, \cdots, p +i_{q-m}\}$. We define $\widetilde{\mu}:\Re^{p+m} \rightarrow \Re$ as the restriction of $\mu : \Re^{p+q} \rightarrow \Re$ subject to $b_{p+i_{1}} =c_{1},  \cdots,  b_{p+i_{q-m}} = c_{q-m}$. Clearly then $\mu(\widetilde{\bb}) = \widetilde{\mu}(\bb)$.
Given a function $\mu$, the fixed value $\bc$  is chosen in such a way that $\mu(\widetilde{\bb}) = \mu^{(\drop)}(\bb).$
We assume that $\bmu: \Re^{p+q}\rightarrow \Re^{\ell}$ is a function that is $1^{{\rm st}}$ order partially differentiable at $\bbetatrue$.
Note that by definition of $\bc$-augmentation, $\mu(\tbbetak) = \mu^{(\drop)}(\hbbetak)$. For ease of reading, in the subsequent proof, we omit the superscript `$(\drop)$'.

\subsection*{A.2. Proof of Theorem \ref{th:main2}}

From usual regularity conditions on the log-likelihood, it can be shown that $\sqrt{n}\left(\hbbeta_k-\bbetakstar\right) = - \matH^{-1}_{k}\left\{\dfrac{1}{\sqrt{n}}\sum^{n}_{i=1}\telli'(\bbetakstar)\right\} + o_{\bbP}(1)$. For more detail and exact conditions see \cite[Chapter~5]{van00}.

Now by application of Taylor expansion, $\mu(\hbbetak) - \mu(\bbetakstar) = \nabla\mu(\bbetakstar)^{\top}(\hbbetak-\bbetakstar) + o_{\bbP}(\norm{\hbbetak-\bbetakstar}{})$, so that
\[
\sqrt{n}(\mu(\hbbetak) - \mu(\bbetakstar)) =  -\nabla\mu(\bbetakstar)^{\top}\left[\matH^{-1}_{k} \left\{\frac{1}{\sqrt{n}}\sum^{n}_{i=1}\telli'(\bbetakstar)\right\} + o_{\bbP}(1)\right] + o_{\bbP}(\sqrt{n}\norm{\hbbetak-\bbetakstar}{}).
\]
Thus it follows that for $0\leq w_{k}\leq 1$ with $\sum_{k\in \calM}w_{k}=1$,
\begin{align*}
& \sqrt{n}\sum_{k \in \calM} w_{k}\{\mu(\hbbetak) - \mu(\bbetatrue)\} \\
& =\sqrt{n} \sum_{k \in \calM} w_{k}\{\mu(\bbetakstar) - \mu(\bbetatrue)\}+  \sqrt{n}\sum_{k \in \calM} w_{k}\{\mu(\hbbetak) - \mu(\bbetakstar)\}\\
& = \sqrt{n} \sum_{k \in \calM} w_{k}\{\mu(\bbetakstar) - \mu(\bbetatrue)\} - \sum_{k \in \calM} w_{k}\nabla\mu(\bbetakstar)^{\top}\matH^{-1}_{k} \left\{\frac{1}{\sqrt{n}}\sum^{n}_{i=1}\telli'(\bbetakstar)\right\} \\
& \hspace{2cm} + o_{\bbP}\left(\sum_{k \in \calM}\sqrt{n}\norm{\hbbetak-\bbetakstar}{}\right)\\
& = \sqrt{n} \sum_{k \in \calM} w_{k}\{\mu(\bbetakstar) - \mu(\bbetatrue)\}+ \dfrac{1}{\sqrt{n}} \sum^{n}_{i=1} \left\{-\sum_{k\in \calM}w_{k}\nabla\mu(\bbetakstar)^{\top}\matH_{k}^{-1}\ellki'(\bbetakstar)\right\} \\
& \hspace{2cm}+ o_{\bbP}\left(\sum_{k \in \calM}\sqrt{n}\norm{\hbbetak-\bbetakstar}{}\right)\\
& = \sqrt{n} \sum_{k \in \calM} w_{k}\{\mu(\bbetakstar) - \mu(\bbetatrue)\}+ \dfrac{1}{\sqrt{n}} \sum^{n}_{i=1} Z_{i} + o_{\bbP}\left(\sum_{k \in \calM}\sqrt{n}\norm{\hbbetak-\bbetakstar}{}\right),
\end{align*}
where we have used the definition that $Z_{i} = -\sum_{k\in \calM}w_{k}\nabla\mu(\bbetakstar)^{\top}\matH_{k}^{-1}\ellki'(\bbetakstar)$. First note that $\sqrt{n}\norm{\hbbetak-\bbetakstar}{} =o_{\bbP}(1)$ via consistency of MLE. Note that $Z_{i}$'s are independent and $\bbE Z_{i} = 0$. Now fix $\epsa>0$. In order to prove the asymptotic normality of the quantity $(1/\sqrt{n})\sum_{i}Z_{i}$ we invoke the Lindeberg-Feller central limit theorem (see \cite{billing08}). This requires verification of the so called Lindeberg condition, given by $(1/n)\sum^{n}_{i=1}\bbE Z^{2}_{i}\bbI\left\{|Z_{i}|>\sqrt{n}\epsa\right\}$. Let us denote $Y_{ki}=\nabla\mu(\bbetakstar)\matH_{k}^{-1}\ellki'$. Now,
\begin{align*}
\dfrac{1}{n}\sum^{n}_{i=1}\bbE Z^{2}_{i}\bbI\left\{|Z_{i}|>\sqrt{n}\epsa\right\} & = \dfrac{1}{n}\sum^{n}_{i=1}\bbE\underbrace{\left(\sum_{k\in \calM}w_{k}Y_{ki}\right)^{2}}_{= A, \text{ say}}\ \underbrace{\bbI\left\{|\sum_{k\in \calM}w_{k}Y_{ki}|>\sqrt{n}\epsa\right\}}_{=B, \text{ say}}\\
& \leq \dfrac{1}{n}\sum^{n}_{i=1}\bbE \left[\sum_{k \in \calM}w_{k}Y_{ki}^{2}\ \bbI\left\{\max_{k \in \calM} |Y_{ki}|>\sqrt{n}\epsa\right\}\right]\\
& \leq \dfrac{1}{n}\sum^{n}_{i=1} \bbE \left[\max_{k \in \calM} |Y_{ki}|^{2}\bbI \ \left\{\max_{k \in \calM} |Y_{ki}|>\sqrt{n}\epsa\right\}\right].
\end{align*}
Here the inequality in the second line is derived by first noting that if $A,B>0$ and $A<C, B<D$, then $AB<CD$. Secondly, note that $A= (\sum_{k \in \calM}w_{k}Y_{ki}) \leq \sum_{k \in \calM}w_{k}Y^{2}_{ki}$ by Jensen's inequality. Also since $\sqrt{n}\epsa < |\sum_{k \in \calM}w_{k}Y_{ki}| \leq \max_{k \in \calM}\sum_{k}|w_{k}| = 1$, it follows that $\bbI\left\{|\sum_{k\in \calM}w_{k}Y_{ki}|>\sqrt{n}\epsa\right\}\leq \bbI\left\{\max_{k \in \calM} |Y_{ki}|>\sqrt{n}\epsa\right\}$. Now take $C=\sum_{k \in \calM}w_{k}Y^{2}_{ki}$ and $D = \bbI\left\{\max_{k \in \calM} |Y_{ki}|>\sqrt{n}\epsa\right\}$.

Now by condition (\textbf{A}1), the Lindeberg-Feller condition is satisfied for $(1/\sqrt{n})Z_{i}$'s whence it follows that $(1/\sqrt{n})\sum^{n}_{i=1}Z_{i} \sim \calN(0, \sigma_{w}^{2})$, where $\sigma^{2}_{w}$ is given by
\[
\sigma^{2}_{w} = \lim_{n\rightarrow \infty} \dfrac{1}{n}\sum^{n}_{i =1} \bbE \left\{\sum_{k}w_{k}\nabla\mu(\bbetakstar)^{\top}\matH^{-1}_{k}\ellki'\right\}^{2}.
\]
The theorem follows.

\subsection*{A.3. Proof of Corollary \ref{cor:match}}

As defined before, for the $k^{{\rm th}}$ candidate model, let
$\bbetakstar \in \Re^{p+|M_{k}|}$ be the solution of the equation $\bbE S_{k}(\bbeta)=0$, where $S_{k}(\bbeta)$ is the score function for the $k^{{\rm th}}$ model. Let $\bbeta_{0,k}= (\btheta_{0}, \pi_{k}\bgamma_{0})^{\top}\in \Re^{p+|M_{k}|} $. Therefore,
$\bbE\{\tell'(\bbetakstar)\}=\bzero$. Then, by Taylor's theorem and appropriate regularity conditions on the density function, it follows that asymptotically, $\bbetakstar - \bbeta_{0,k} \approx \matJ_{k}^{-1}\bbE\{\tell'(\bbeta_{0})\}$. Now note that following \cite[Page~37]{Hjort2003},
\[
\bbE\{\tell'(\bbeta_{0})\} = \begin{pmatrix}
\matJ_{01}\bdelta/\sqrt{n} + o(1/\sqrt{n}) \\ \pi_{k}\matJ_{11}\bdelta/\sqrt{n} + o(1/\sqrt{n})
\end{pmatrix},
\]
so that,
\begin{align}\label{eq:imp1}
\bbetakstar - \bbeta_{0,k} \approx \matJ_{k}^{-1} \begin{pmatrix}
\matJ_{01}\bdelta/\sqrt{n}  \\ \pi_{k}\matJ_{11}\bdelta/\sqrt{n} \end{pmatrix}.
\end{align}
In order to prove the corollary, we first match the bias terms. Note that in Theorem \ref{th:main2}, the bias term is given by
\[
\sqrt{n} \sum_{k \in \calM} w_{k}\{\mu(\bbetakstar, \gamma_{0,k^{c}}) - \mu(\bbetatrue)\}.
\]
Thus consider term by term, the bias of the $k^{\rm th}$ component is given by
\begin{align*}
\sqrt{n}\{\mu(\bbetakstar, \bgamma_{0,k^{c}}) - \mu(\bbetatrue)\}& = \sqrt{n}\{\mu(\bbetakstar, \bgamma_{0,k^{c}}) - \mu(\bbeta_{0})\}- \sqrt{n}\{\mu(\bbetatrue)- \mu(\bbeta_{0})\}\\
& \approx \sqrt{n}(\bbetakstar-\bbeta_{0,k})^{\top}\begin{pmatrix}
\pa\mu(\bbeta_{0})/\pa \btheta \\ \pa\mu(\bbeta_{0})/\pa \bgamma_{k}
\end{pmatrix} - \bigg(\dfrac{\pa\mu(\bbeta_{0})}{\pa \bgamma}\bigg)^{\top}\bdelta\\
& = \begin{pmatrix}
\pa\mu(\bbeta_{0})/\pa \btheta \\ \pa\mu(\bbeta_{0})/\pa \bgamma_{k}
\end{pmatrix}^{\top} \matJ_{k}^{-1} \begin{pmatrix}
\matJ_{01}\bdelta \\ \pi_{k}\matJ_{11}\bdelta \end{pmatrix} -  \bigg(\dfrac{\pa\mu(\bbeta_{0})}{\pa \bgamma}\bigg)^{\top}\bdelta,
\end{align*}
where the last term follows from (\ref{eq:imp1}). This matches the bias term in (\ref{eq:hjortasymp}). Looking at the variance term, note that from (\ref{eq:varmu}), the variance of the $k^{\rm th}$ term is given by,

\[
\mbox{\rm var}\{\nabla\mu(\bbetakstar, \bgamma_{0,k^{c}})\}^{\top}\matH^{-1}_{k}(\sum^{n}_{i=1}\ellki'(\bbetakstar)/\sqrt{n}).
\]
From (\ref{eq:imp1}), via Taylors theorem it follows that $\nabla\mu(\bbetakstar, \bgamma_{0,k^{c}}) \approx \nabla\mu(\bbeta_{0})$. Also note that from standard theory of maximum likelihood estimation,

\begin{align*}
\matH^{-1}_{k}(\bbetakstar)(\sum^{n}_{i=1}\ellki'(\bbetakstar)/\sqrt{n}) & \approx \sqrt{n} (\hbbetak-\bbetakstar)\\
& = \sqrt{n} (\hbbetak-\bbeta_{0,k}) - \sqrt{n} (\bbetakstar - \bbeta_{0,k})\\
& = \matJ_{k}^{-1} \begin{pmatrix}
\sqrt{n}\Ubar_{n}\\ \sqrt{n}\Vbar_{n,k}
\end{pmatrix} - \matJ_{k}^{-1} \begin{pmatrix}
\matJ_{01}\bdelta \\ \pi_{k}\matJ_{11}\bdelta \end{pmatrix}\\
& = \matJ_{k}^{-1}\begin{pmatrix}
\sqrt{n}\{\Ubar_{n}- \bbE U_{k}(Y_{1})\}\\ \sqrt{n}\{\Vbar_{n,k} - \bbE V_{k}(Y_{1})\}
\end{pmatrix}.
\end{align*}
Here the last inequality follows from Lemma 3.1 in \cite{Hjort2003}. Hence it follows that asymptotically both the bias and variance terms are equal.




\bibliography{paper4thnov}
\end{document}